\documentclass[fleqn,a4paper]{article}

\usepackage[style=ieee,sorting=none,maxbibnames=99]{biblatex}

\usepackage{amsmath,amssymb,amsthm,amsfonts}
\usepackage{hyperref}
\usepackage[margin=2cm]{geometry}
\usepackage{graphicx}
\usepackage[curve,matrix,arrow,color]{xy}
\usepackage{tikz}
\usepackage{verbatim}
\usepackage{tikz-cd}
\usetikzlibrary{calc}
\usetikzlibrary{decorations.pathreplacing}
\usetikzlibrary{decorations.markings}
\usetikzlibrary{intersections} 
\usetikzlibrary{er,positioning}
\usetikzlibrary{arrows,shapes}
\usetikzlibrary{decorations.markings}

\usepackage{booktabs}
\usepackage{multirow}
\usepackage{makecell}

\newcommand{\e}{\mathrm e}
\renewcommand{\i}{\mathrm i}
\DeclareMathOperator{\sgn}{sgn}
\DeclareMathOperator{\tr}{tr}

\usepackage[asterism]{sectionbreak}
\usepackage{xparse}
\DeclareDocumentCommand\ket{ s m }
{ 
	\IfBooleanTF{#1}
	{\lvert{#2}\rangle} 
	{\left\lvert{#2}\right\rangle} 
}

\DeclareDocumentCommand\innerproduct{ s m g }
{ 
	\IfBooleanTF{#1}
	{ 
		\IfNoValueTF{#3}
		{\langle{#2}\vert{#2}\rangle}
		{\langle{#2}\vert{#3}\rangle}
	}
	{ 
		\IfNoValueTF{#3}
		{\left\langle{#2}\middle\vert{#2}\right\rangle}
		{\left\langle{#2}\middle\vert{#3}\right\rangle}
	}
}
\DeclareDocumentCommand\braket{}{\innerproduct} 
\DeclareDocumentCommand\ip{}{\innerproduct} 
\DeclareDocumentCommand\expectationvalue{ s s m g }
{ 
	\IfNoValueTF{#4}
	{
		\IfBooleanTF{#1}
		{\langle{#3}\rangle} 
		{\left\langle{#3}\right\rangle} 
	}
	{
		\IfBooleanTF{#1}
		{
			\IfBooleanTF{#2}
			{\left\langle{#4}\middle\vert{#3}\middle\vert{#4}\right\rangle} 
			{\langle{#4}\vert{#3}\vert{#4}\rangle} 
		}
		{\left\langle{#4}\middle\vert\smash{#3}\middle\vert{#4}\right\rangle} 
	}
}

\DeclareDocumentCommand\matrixelement{ s s m m m }
{ 
	\IfBooleanTF{#1}
	{
		\IfBooleanTF{#2}
		{\left\langle{#3}\middle\vert{#4}\middle\vert{#5}\right\rangle} 
		{\langle{#3}\vert{#4}\vert{#5}\rangle} 
	}
	{\left\langle{#3}\middle\vert\smash{#4}\middle\vert{#5}\right\rangle} 
}

\DeclareDocumentCommand\mel{}{\matrixelement} 

\DeclareDocumentCommand\ev{}{\expectationvalue}

\DeclareDocumentCommand\qqtext{ s m }{\IfBooleanTF{#1}{}{\quad}\text{#2}\quad}
\DeclareDocumentCommand\qq{}{\qqtext}

\DeclareDocumentCommand\partialderivative{ s o m g g d() }
{ 
	\IfBooleanTF{#1}
	{\let\fractype\flatfrac}
	{\let\fractype\frac}
	\IfNoValueTF{#4}
	{
		\IfNoValueTF{#6}
		{\fractype{\partial \IfNoValueTF{#2}{}{^{#2}}}{\partial #3\IfNoValueTF{#2}{}{^{#2}}}}
		{\fractype{\partial \IfNoValueTF{#2}{}{^{#2}}}{\partial #3\IfNoValueTF{#2}{}{^{#2}}} \argopen(#6\argclose)}
	}
	{
		\IfNoValueTF{#5}
		{\fractype{\partial \IfNoValueTF{#2}{}{^{#2}} #3}{\partial #4\IfNoValueTF{#2}{}{^{#2}}}}
		{\fractype{\partial^2 #3}{\partial #4 \partial #5}}
	}
}

\newcommand{\overbar}[1]{\mkern 1.5mu\overline{\mkern-1.5mu#1\mkern-1.5mu}\mkern 1.5mu}
\usepackage{newtxtext,newtxmath}

\newcommand{\q}{\mathfrak{q}}
\newcommand{\m}{\mathfrak{m}}

\newcommand{\JTL}[1]{\mathsf{JTL}_{#1}}
\newcommand{\ATL}[1]{\mathsf{T}^{\mathrm a}_{#1}}

\renewcommand{\i}{\mathrm i}
\newcommand{\Verma}[1]{\mathsf{V}_{#1}}
\newcommand{\Vermab}[1] {\overbar{\mathsf{V}}_{#1}}
\newcommand{\IrrV}[1]{\mathsf{X}_{#1}}
\newcommand{\IrrVb}[1]{\overbar{\mathsf{X}}_{#1}}
\newcommand{\Vir}{\hbox{Vir}}
\newtheorem{conjecture}{Conjecture}
\newcommand{\loopinner}[2]{\langle #1, #2 \rangle}

\newcommand{\loopdagger}{\dagger}

\newcommand{\KSL}{L^{\tiny{(L)}}}
\newcommand{\KSLb}{\overbar{L}^{\tiny{(L)}}}

\title{Emerging Jordan blocks in the two-dimensional Potts and loop models at generic $Q$}

\author{Lawrence Liu$^1$, Jesper Lykke Jacobsen$^{2,3,4}$, Hubert Saleur$^{1,2}$ \\
[2.0mm]
${}^1$ \small Department of Physics, University of Southern California, Los Angeles, CA 90089-0484, USA\\
${}^2$ \small Universit\'e Paris-Saclay, CNRS, CEA, Institut de Physique Th\'eorique, 91191, Gif-sur-Yvette, France \\
${}^3$ \small Laboratoire de Physique de l'\'Ecole Normale Sup\'erieure, ENS, Universit\'e PSL, CNRS, Sorbonne Universit\'e, Universit\'e de Paris\\ 
${}^4$ \small Sorbonne Universit\'e, \'Ecole Normale Sup\'erieure, CNRS, Laboratoire de Physique (LPENS), 75005 Paris, France}

\date{28 March 2024}

\begin{document}

\maketitle

\begin{abstract}
It was recently suggested---based on general self-consistency arguments as well as results from the bootstrap \cite{GorbenkoZan2020,Grans-Samuelsson2020,NR_2021}---that the CFT describing the $Q$-state Potts model is logarithmic for generic values of $Q$, with rank-two Jordan blocks for $L_0$ and $\overbar{L}_0$ in many sectors of the theory. This is despite the well-known fact that the lattice transfer matrix (or Hamiltonian) is diagonalizable in (arbitrary) finite size. While the emergence of Jordan blocks only in the limit $L\to\infty$ is perfectly possible conceptually, diagonalizability in finite size makes the measurement of logarithmic couplings (whose values are analytically predicted in \cite{Grans-Samuelsson2020,NR_2021}) very challenging. This problem is solved in the present paper (which can be considered a companion to \cite{Grans-Samuelsson2020}), and the conjectured logarithmic structure of the CFT confirmed in detail by the study of the lattice model and associated ``emerging Jordan blocks.''

\end{abstract}

\section{Introduction}

The bootstrap approach and the detailed study of four-point correlation functions have helped clarify the logarithmic properties of the $Q$-state Potts model and $O(\m)$ model conformal field theories (CFTs) for generic values of the parameters $Q$ and $\m$ (see, e.g., the special volume \cite{Gainutdinov_2013} for a general review of the topic of logarithmic CFTs). Many of the results in this field, however, rely on self-consistency arguments, and have not been checked directly, particularly via numerical calculations. A noticeable exception is the $c = 0$ case, where the Jordan block mixing the stress--energy tensor $T$ and its logarithmic partner $t$ \cite{Gurarie1993,GurarieLudwig2002} was first observed in \cite{DJS2010c}, leading to measurements of the famous ``$b$-numbers''---the logarithmic couplings---and the resolution of several paradoxes involving the polymer--percolation and bulk--boundary differences \cite{VJS2011,VGJS2012}. This progress at $c = 0$ was rendered considerably easier by the presence, already in finite size, of Jordan blocks for the Hamiltonian going over to the CFT Jordan blocks for $L_0$ and $\overbar L_0$ in the continuum limit.

By contrast, despite the conjectured appearance of Jordan blocks in $L_0$ and $\overbar L_0$ for the $Q$-state Potts and $O(\m)$ models in the continuum limit \cite{GorbenkoZan2020,Grans-Samuelsson2020}, their known lattice versions have Hamiltonians which, although non-hermitian, are fully diagonalizable for generic values of $Q$ and $\m$, making the confirmation of their logarithmic structures---let alone the measurement of the corresponding logarithmic couplings---significantly more difficult, both conceptually and technically.

Our goal in this paper is to surmount this obstacle by devising strategies to study ``emerging'' Jordan blocks and thus confirm, in particular, the predictions of \cite{Grans-Samuelsson2020,NR_2021}. While we focus here on a specific lattice model---the oriented loop model in the terminology of
\cite{ReadSaleur2007} (this model is also referred to as the $U(\m)$ model in \cite{JRS2022}) describing the low-temperature $O(\m)$ model as well as hulls of $Q$-state Potts model clusters\footnote{As discussed in detail in \cite{JRS2022}, there is a small but significant difference between the low temperature phase of the $O(\m)$ model and the $U(\m)$ model having to do with the allowed parities of non-contractible lines. This does not affect any of the results of this paper.}---the proposed approach is completely general.

\sectionbreak

Since this paper comes on the heels of many works devoted to the same lattice models and associated CFTs, we have largely suppressed reminders of well-established facts, and follow as closely as possible the notations of our previous papers on the subject, especially references \cite{Grans-Samuelsson2020,GSJS2021}. There is a very important exception to this, however: in \cite{Grans-Samuelsson2020,GSJS2021} we used the notation $(\cdot|\cdot)$ for the loop scalar product and $\ip{\cdot}$ for the natural, positive-definite scalar product. For the present paper we have decided to switch notations, and thus represent the loop scalar product by $\ip{\cdot}$, so that it is denoted by the same symbol as the product ordinarily used in CFT to represent, for example, correlation functions.

The remainder of this paper is organized as follows. Section \ref{preliminaries} establishes the context of our present study as well as our notations. Though much of this material appears in some form in our previous work on the subject \cite{Grans-Samuelsson2020,GSJS2021}, there are some important additions and revisions that clarify the exposition, besides the shift in notation for inner products already mentioned. In Section \ref{sec:TL}, we give a new exposition of the representation theory of Temperley--Lieb algebras, showing how the standard modules relate to what we call natural representations, and clarify how the nature of the representations changes when we continue to modify the Temperley--Lieb algebras. We also introduce a notation for link states, which form the bases of our representations. In Section \ref{conjectures} we repeat our conjecture whose verification is a primary aim of this paper. In Section \ref{strategy} we introduce our methods and techniques for verifying this conjecture, as well as the difficulties that necessitated the introduction of these techniques. Section \ref{measurements} then shows the results of our calculations and discusses their interpretation, and Section \ref{conclusion} concludes. The appendix explains in detail, with numerical examples, some subtleties present in the model we study that had not previously arisen, and uses our new emerging Jordan blocks technique to replicate earlier results, thereby enhancing its reliability.

\subsection*{Notations and definitions}

We gather here some general notations and definitions that are used throughout the paper:

\begin{itemize}

\item $\ATL{N}(\m)$ --- the affine Temperley--Lieb algebra on $N$ sites with parameter $\m$. We shall parametrize the loop weight as $\m=\q+\q^{-1}$, with $\q=\e^{\i\gamma}$ and $\gamma=\pi/(x+1)$ with $x > 0$. The term ``generic'' indicates $x \in \mathbb{R}^+ \setminus \mathbb{Q}$---that is, $x$ is irrational, or equivalently, $\q$ is not a root of unity.

\item $\mathscr W_{j,\e^{2\pi\i K}}$ --- standard module of the affine Temperley--Lieb algebra with $2j$ through-lines and pseudomomentum $K$.

\item $\overbar{\mathscr W}_{\!\!0,\mathfrak q^{\pm 2}}$ --- the irreducible quotient of the standard module $\mathscr W_{0,\q^{\pm 2}}$.

\item $\mathscr N_j$ --- the natural representation of $\ATL{N}(\m)$ consisting of all link states with $2j$ or fewer through-lines and no pseudomomentum. It can be viewed as the vector space sum of $\mathscr W_{0,\q^{\pm 2}}$ and all $\mathscr W_{k,1}$ with $1 \le k \le j$, with the module structure given by the natural action of the diagrams.

\item $\overbar{\mathscr N}_{\!\!j}$ --- a quotient of $\mathscr N_j$ that essentially replaces $\mathscr W_{0,\q^{\pm 2}}$ with $\overbar{\mathscr W}_{\!\!0,\mathfrak q^{\pm 2}}$.

\item $\Verma{r,s}$ --- Verma module for the conformal weight $h_{r,s}$, when either $r \notin \mathbb{N}^*$ or $s \notin \mathbb{N}^*$.

\item $\Verma{r,s}^{\rm d}$ --- the (degenerate) Verma module for the conformal weight $h_{r,s}$, when $r,s\in\mathbb{N}^*$.

\item $\IrrV{r,s}$ --- irreducible Virasoro module for the conformal weight $h_{r,s}$.

\item{} A conformal weight $h_{r,s}$ with $r,s\in\mathbb{N}^*$ will be called degenerate. For such a weight, 
there exists a descendant state that is also primary: this descendant is often called a null (or singular) vector (or state). We will denote by $A_{r,s}$ the combination of Virasoro generators producing the null state at level $r s$ corresponding to the degenerate weight $h_{r,s}$. $A_{r,s}$ is normalized so that the coefficient of $L_{-rs}$ is equal to unity. Some examples are
\begin{subequations}
\label{A_operators}
\begin{gather}
A_{1,1} = L_{-1}, \\
A_{1,2} = L_{-2}-\frac{3}{2(2h_{1,2}+1)}L_{-1}^2, \qq{and}\label{A_12} \\
A_{2,1} = L_{-2}-\frac{3}{2(2h_{2,1}+1)}L_{-1}^2.
\end{gather}
\end{subequations}

\item{} In the bulk of the paper we consider only generic values of the parameter $\q$ (i.e., $\q$ not a root of unity), and thus generic values of $x$ (i.e., $x$ irrational). It can happen, however, that the modules of interest are not irreducible even then. We call them ``non-generic'' in the affine Temperley--Lieb case, and ``degenerate'' in the Virasoro case. In earlier papers (see, e.g., \cite{GJS2018}), we have used the names ``partly non-generic'' and ``partly degenerate,'' respectively, since having $\q$ a root of unity adds considerably more structure to the modules. We will not do so here, the context clearly excluding $\q$ a root of unity---except for the appendix. 

\item{} The continuum limit of the loop scalar product on the lattice is the usual Virasoro scalar product, for which we use $\dagger$ to denote conjugates so $\loopinner{V_1}{L_n V_2} = \loopinner{L_n^\loopdagger V_1}{V_2}$. When using this scalar product we shall also use the bra-ket notation: $|V\rangle$ denotes a state $V$ (primary or not) and $\langle V|$ its dual.

\item{} We denote by $\phi_{r,s}$ a chiral primary field with conformal weight $h_{r,s}$ and $r,s\in\mathbb{R}$: the structure of the underlying Virasoro module when $r,s\in \mathbb{N}^*$ will be made clear from the context, but will not appear in the notation. We will also freely make use of the symmetries $h_{r,s}=h_{-r,-s}$. 

\end{itemize}

\section{Preliminaries} \label{preliminaries}
\subsection{From the Potts model to the dense loop model and its algebraic description} \label{loop_model}

Sections 2 and 3 of \cite{Grans-Samuelsson2020} contain background material establishing the correspondence between the configurations of Fortuin--Kasteleyn clusters (which might be more familiar to the reader) and those of loops. It is somewhat more convenient to use the loop formalism, which makes the relationship with the $O(\m)$ model more transparent from the beginning, and connects more naturally with work on diagram algebras. The remainder of this section describes the various algebraic aspects of the loop model in detail as well as the representations to be studied.

\subsection{Representations of various Temperley--Lieb algebras} \label{sec:TL}

\subsubsection{Temperley--Lieb algebras}

In terms of generators and relations, various algebras to which the name ``Temperley--Lieb'' (TL) is associated begin with generators $\{e_j\}$ subject to relations
\begin{subequations} \label{eq:TL_relations}
\begin{gather}
e^2_j = \m e_j, \label{eq:TL_m}\\
e_j e_{j\pm 1}e_j = e_j, \qq{and} \\
e_j e_k = e_k e_j, \qquad (|j - k| > 1)
\end{gather}
\end{subequations}
where the indices take the values $1,\ldots,N-1$, with $N$ a fixed integer, and each value of $N$ defines a different algebra. These relations are typically illustrated using diagrams for the generators. To represent generator $e_j$, on two horizontal rows of $N$ points each, connect the neighboring pairs at $j$ and $j+1$, and vertically match all other pairs:
\begin{equation} \label{eq:e_j_diagram}
e_j =
\raisebox{-5mm}{
\begin{tikzpicture}[scale=0.6]
 \draw[thick] (0,-1)--(0,1) node[above]{\scriptsize $1$};
\end{tikzpicture}}
\quad\cdots\quad
\raisebox{-5mm}{\begin{tikzpicture}[scale=0.6]
 \draw[thick] (-1,-1)--(-1,1);
 \draw[thick] (2,-1)--(2,1);
 \draw[thick] (0,-1)--(0,-0.5) arc(180:0:5mm and 4mm)--(1,-1);
 \draw[thick] (0,1) node[above]{\scriptsize $j$}--(0,0.5) arc(180:360:5mm and 4mm)--(1,1) node[above]{\scriptsize $j+1$};
\end{tikzpicture}}
\quad\cdots\quad
\raisebox{-5mm}{\begin{tikzpicture}[scale=0.6]
 \draw[thick] (0,-1)--(0,1) node[above]{\scriptsize $N$};
\end{tikzpicture}}
\end{equation}
When two generators are multiplied, the corresponding diagram is obtained following certain rules: stack the diagrams from top to bottom (corresponding to left to right multiplication), replace closed loops with a numerical factor $m$, and straighten out the curves that connect vertically. In any diagram of the algebra, the lines that connect from top to bottom (rather than between two sites in the same horizontal row) are called \emph{through-lines}, and the sites to which they are connected are called \emph{free}. The identity element $1$ consists of a diagram with all $N$ vertical pairs of sites connected by straight through-lines. It can be shown formally that this algebra of diagrams is isomorphic to the abstract Temperley--Lieb algebra defined by generators and relations \cite{RidoutSaintAubin2014}. A basis for this algebra is the set of all isotopically distinct diagrams---pairings of two rows of $N$ points each without crossings---that can be obtained by repeated multiplication of the generators.

A simple generalization is to adjoin a generator $e_N$ and site $N+1$, and to identify site $N+1$ with site $1$, thereby obtaining a periodic Temperley--Lieb algebra. The $N$ sites in the top and bottom rows can be thought of as being placed on the inner and outer boundaries of an annulus, evenly spaced, and multiplication consists of nesting two annuli after rescaling one so that its inner boundary matches the outer boundary of the other. In both cases the accessible diagrams consist of planar pairings of points using curves that do not intersect. As before, the curves that join points on the inner boundary to points on the outer boundary of the annulus are called through-lines. We may also depict the multiplication more conveniently using ``framing'' rectangles, whose left and right boundaries are identified. Such framing rectangles will be shown using dotted boundaries.

For the periodic case, it is possible for noncontractible closed curves to form that circle the annulus (this requires $N$ to be even). These will be given the same weight $\m$ as contractible loops, although one may consider giving contractible and noncontractible loops different weights---with interesting consequences.
 
Some important representations of the Temperley--Lieb and periodic Temperley--Lieb algebras are obtained from the following considerations. Begin with the regular representation $R$ on the basis given by the left action: $R_a b = ab$, where $a$ and $b$ are basis diagrams. Since left-to-right multiplication corresponds to top-to-bottom stacking of diagrams, the pairings in the bottom row of the composite diagram $ab$ will be the same as those of the bottom row of $b$---the left multiplication by $a$ will connect to only the top row of $b$. The output of the multiplication by $a$ is therefore largely determined by the top row of $b$. We may therefore ``cut across'' the through-lines of the basis diagrams and retain only the top halves, which are single rows of $N$ points, some of which are joined by curves and others that remain free. The lines attached to free sites are also called through-lines---in the transfer matrix picture they may be thought of as being connected to the remote past.

We illustrate this process explicitly for the periodic Temperley--Lieb algebra on $N = 2$ sites. The generators and the identity element are
\begin{equation} \label{eq:e_j_diagram}
e_1 = \;\raisebox{-5mm}{\begin{tikzpicture}[scale=0.6]
 \draw[thick] (0,-1)--(0,-0.5) arc(180:0:5mm and 4mm)--(1,-1);
 \draw[thick] (0,1)--(0,0.5) arc(180:360:5mm and 4mm)--(1,1);
 \draw[dotted] ($(current bounding box.south east)+(5mm,0)$) rectangle ($(current bounding box.north west)-(5mm,0)$);
\end{tikzpicture}},\quad
e_2 = \;\raisebox{-5mm}{\begin{tikzpicture}[scale=0.6]
 \draw[thick] (0.5,-1)--(0.5,-0.5) arc(0:90:5mm and 4mm);
 \draw[thick] (0.5,1)--(0.5,0.5) arc(0:-90:5mm and 4mm);
 \draw[thick] (1.5,-1)--(1.5,-0.5) arc(180:90:5mm and 4mm);
 \draw[thick] (1.5,1)--(1.5,0.5) arc(180:270:5mm and 4mm);
 \draw[dotted] ($(current bounding box.south east)$) rectangle ($(current bounding box.north west)$);
\end{tikzpicture}},\quad
\mathrm{id} = \;\raisebox{-5mm}{
\begin{tikzpicture}[scale=0.6]
 \draw[thick] (0,-1)--(0,1);
 \draw[thick] (1,-1)--(1,1);
 \draw[dotted] ($(current bounding box.south east)+(5mm,0)$) rectangle ($(current bounding box.north west)-(5mm,0)$);
\end{tikzpicture}}
\end{equation}
(the dotted borders serve as a reminder that the left and right boundaries are identified). Further multiplication of the diagrams gives
\begin{equation} \label{eq:e_j_diagram}
e_1e_2 = \m\times{}
\;\raisebox{-5mm}{\begin{tikzpicture}[scale=0.6]
 \draw[thick] (0.5,-1)--(0.5,-0.5) arc(0:90:5mm and 4mm);
 \draw[thick] (0.5,1)--(0.5,0.5) arc(180:360:5mm and 4mm)--(1.5,1);
 \draw[thick] (1.5,-1)--(1.5,-0.5) arc(180:90:5mm and 4mm);
 \draw[dotted] ($(current bounding box.south east)$) rectangle ($(current bounding box.north west)$);
\end{tikzpicture}},\quad
e_2e_1 = \m\times{}
\;\raisebox{-5mm}{\begin{tikzpicture}[scale=0.6]
 \draw[thick] (0.5,-1)--(0.5,-0.5) arc(180:0:5mm and 4mm)--(1.5,-1);
 \draw[thick] (0.5,1)--(0.5,0.5) arc(0:-90:5mm and 4mm);
 \draw[thick] (1.5,1)--(1.5,0.5) arc(180:270:5mm and 4mm);
 \draw[dotted] ($(current bounding box.south east)$) rectangle ($(current bounding box.north west)$);
\end{tikzpicture}}.
\end{equation}
The distinct upper halves are then \ \begin{tikzpicture}[scale=0.3]
 \draw[thick] (0.5,1)--(0.5,0.5) arc(180:360:5mm and 4mm)--(1.5,1);
 \draw[dotted] ($(current bounding box.south east)+(5mm,0)$) rectangle ($(current bounding box.north west)-(5mm,0)$);
\end{tikzpicture}, \ \begin{tikzpicture}[scale=0.3]
 \draw[thick] (0.5,1)--(0.5,0.5) arc(0:-90:5mm and 4mm);
 \draw[thick] (1.5,1)--(1.5,0.5) arc(180:270:5mm and 4mm);
 \draw[dotted] ($(current bounding box.south east)$) rectangle ($(current bounding box.north west)$);
\end{tikzpicture}, and \ \begin{tikzpicture}[scale=0.3]
 \draw[thick] (0,0)--(0,1);
 \draw[thick] (1,0)--(1,1);
 \draw[dotted] ($(current bounding box.south east)+(5mm,0)$) rectangle ($(current bounding box.north west)-(5mm,0)$);
\end{tikzpicture}. Similarly, the distinct upper halves of the basis diagrams resulting from the same process applied at $N = 4$ are \ \begin{tikzpicture}[scale=0.3]
 \draw[thick] (0.5,1)--(0.5,0.5) arc(180:360:5mm and 4mm)--(1.5,1);
 \draw[thick] (2.5,1)--(2.5,0.5) arc(180:360:5mm and 4mm)--(3.5,1);
 \draw[dotted] ($(current bounding box.south east)+(5mm,0)$) rectangle ($(current bounding box.north west)-(5mm,0)$);
\end{tikzpicture}, \ \begin{tikzpicture}[scale=0.3]
 \draw[thick] (1.5,1)--(1.5,0.5) arc(180:360:5mm and 4mm)--(2.5,1);
 \draw[thick] (0.5,1)--(0.5,0.5) arc(0:-90:5mm and 4mm);
 \draw[thick] (3.5,1)--(3.5,0.5) arc(180:270:5mm and 4mm);
 \draw[dotted] ($(current bounding box.south east)$) rectangle ($(current bounding box.north west)$);
\end{tikzpicture}, \ \begin{tikzpicture}[scale=0.3]
 \draw[thick] (0.5,1)--(0.5,0.5) arc(180:360:15mm and 4mm)--(3.5,1);
 \draw[thick] (1.5,1)--(1.5,0.7) arc(180:360:5mm and 4mm)--(2.5,1);
 \draw[dotted] ($(current bounding box.south east)+(5mm,0)$) rectangle ($(current bounding box.north west)-(5mm,0)$);
\end{tikzpicture}, \ \begin{tikzpicture}[scale=0.3]
 \draw[thick] (0.5,1)--(0.5,0.5) arc(0:-45:15mm and 4mm);
 \draw[thick] (1.5,1)--(1.5,0.5) arc(180:315:15mm and 4mm);
 \draw[thick] (2.5,1)--(2.5,0.7) arc(180:270:5mm and 4mm);
 \draw[thick] (3.5,1)--(3.5,0.7) arc(0:-90:5mm and 4mm);
 \draw[dotted] ($(current bounding box.south east)$) rectangle ($(current bounding box.north west)$);
\end{tikzpicture}, \ \begin{tikzpicture}[scale=0.3]
 \draw[thick] (1.5,1)--(1.5,0.5) arc(0:-90:15mm and 4mm);
 \draw[thick] (2.5,1)--(2.5,0.5) arc(180:270:15mm and 4mm);
 \draw[thick] (3.5,1)--(3.5,0.7) arc(180:270:5mm and 4mm);
 \draw[thick] (0.5,1)--(0.5,0.7) arc(0:-90:5mm and 4mm);
 \draw[dotted] ($(current bounding box.south east)$) rectangle ($(current bounding box.north west)$);
\end{tikzpicture}, \ \begin{tikzpicture}[scale=0.3]
 \draw[thick] (2.5,1)--(2.5,0.5) arc(0:-135:15mm and 4mm);
 \draw[thick] (3.5,1)--(3.5,0.5) arc(180:225:15mm and 4mm);
 \draw[thick] (0.5,1)--(0.5,0.7) arc(180:270:5mm and 4mm);
 \draw[thick] (1.5,1)--(1.5,0.7) arc(0:-90:5mm and 4mm);
 \draw[dotted] ($(current bounding box.south east)$) rectangle ($(current bounding box.north west)$);
\end{tikzpicture}, \ \begin{tikzpicture}[scale=0.3]
 \draw[thick] (0.5,1)--(0.5,0.5) arc(180:360:5mm and 4mm)--(1.5,1);
 \draw[thick] (2.5,0)--(2.5,1);
 \draw[thick] (3.5,0)--(3.5,1);
 \draw[dotted] ($(current bounding box.south east)+(5mm,0)$) rectangle ($(current bounding box.north west)-(5mm,0)$);
\end{tikzpicture}, \ \begin{tikzpicture}[scale=0.3]
 \draw[thick] (1.5,1)--(1.5,0.5) arc(180:360:5mm and 4mm)--(2.5,1);
 \draw[thick] (0.5,0)--(0.5,1);
 \draw[thick] (3.5,0)--(3.5,1);
 \draw[dotted] ($(current bounding box.south east)+(5mm,0)$) rectangle ($(current bounding box.north west)-(5mm,0)$);
\end{tikzpicture}, \ \begin{tikzpicture}[scale=0.3]
 \draw[thick] (2.5,1)--(2.5,0.5) arc(180:360:5mm and 4mm)--(3.5,1);
 \draw[thick] (0.5,0)--(0.5,1);
 \draw[thick] (1.5,0)--(1.5,1);
 \draw[dotted] ($(current bounding box.south east)+(5mm,0)$) rectangle ($(current bounding box.north west)-(5mm,0)$);
\end{tikzpicture}, \ \begin{tikzpicture}[scale=0.3]
 \draw[thick] (1.5,0)--(1.5,1);
 \draw[thick] (2.5,0)--(2.5,1);
 \draw[thick] (0.5,1)--(0.5,0.5) arc(0:-90:5mm and 4mm);
 \draw[thick] (3.5,1)--(3.5,0.5) arc(180:270:5mm and 4mm);
 \draw[dotted] ($(current bounding box.south east)$) rectangle ($(current bounding box.north west)$);
\end{tikzpicture}, and \ \begin{tikzpicture}[scale=0.3]
 \draw[thick] (0,0)--(0,1);
 \draw[thick] (1,0)--(1,1);
 \draw[thick] (2,0)--(2,1);
 \draw[thick] (3,0)--(3,1);
 \draw[dotted] ($(current bounding box.south east)+(5mm,0)$) rectangle ($(current bounding box.north west)-(5mm,0)$);
\end{tikzpicture}. Such diagrams will be called ``link diagrams'' or ``link states'' hereinafter. The collection of link states on $N$ sites furnishes a basis for the ``natural representation'' of the periodic Temperley--Lieb algebra on $N$ sites. (For the ordinary Temperley--Lieb algebra the link states are restricted to those that do not cross the boundary of the framing rectangle.) The action of a basis diagram of the Temperley--Lieb algebra on a link state results in another link state. The natural representation is a module over its corresponding Temperley--Lieb algebra.

At this point we introduce the following convenient notation for link states, useful both as a compact notation and for computation. A link state is represented by a collection of ordered pairings and singletons. The explicit notation of singletons is optional, and a singleton $(i)$ represents a through-line at site $i$. A pairing $(ij)$ or $(i,j)$ denotes a curve that starts from site $i$ and goes forward to $j$ in the rightward direction. When $i > j$ such a curve starts from site $i$, goes past the last site $N$ and crosses the boundary, returns periodically from behind site $1$, then terminates at site $j$. For the three link states at $N = 2$ exhibited graphically, their representations in this notation are $(12)$, $(21)$, and $(1)(2)$. For the $N = 4$ states, they are $(12)(34)$, $(23)(41)$, $(14)(23)$, $(21)(34)$, $(32)(41)$, $(43)(12)$, $(12)(3)(4)$, $(23)(1)(4)$, $(34)(1)(2)$, $(41)(2)(3)$, and $(1)(2)(3)(4)$. Since singletons are optional, these last five may well be denoted $(12)$, $(23)$, $(34)$, $(41)$, and $()$. Of course, $N$ must be understood from context to avoid ambiguities, such as the two instances of $(12)$ in $N = 2$ and $4$, but exactly the same ambiguity arises in the representations of elements of the symmetric group $S_N$ by disjoint cycles when fixed points are omitted. And, as with $S_N$, the order of the pairings and singletons is immaterial, though the order of sites within a pairing is important. In this notation, the action of $e_j$ on a state with singletons $(j)(j+1)$ is $e_j(j)(j+1)[\cdots] = (j,j+1)[\cdots]$, where $[\cdots]$ stands for all other pairs, and $e_j(j,j+1)[\cdots] = \m(j,j+1)[\cdots]$. For purposes of computation, the action of $e_j$ on a general state can be reduced to an examination of several cases.

Hereinafter, we restrict to values of $N$ that are even. For even $N$, the number of paired sites is even for a given link state, and thus the number of through-lines is also even, a number we parametrize as $2j$. The action of the Temperley--Lieb algebra on a link state can only keep constant or reduce the number of through-lines (this only needs to be verified on the generators $e_i$). The natural representation thus admits a filtration by submodules:
\begin{equation}
\mathscr N_{N/2} \supset \mathscr N_{N/2 - 1} \supset \cdots \supset \mathscr N_j \supset \cdots \supset \mathscr N_0 \supset 0.
\end{equation}
The notation $\mathscr N_j$ indicates the submodule of all link states with $2j$ or fewer through-lines; we will consider these as well to be natural representations. The factors of this filtration are the subquotients $\mathscr W_{j,1} \equiv \mathscr N_j/\mathscr N_{j-1}$ and consist of link diagrams with exactly $2j$ through-lines. The final factor is denoted $\mathscr W_{0,\q^{\pm 2}} \equiv \mathscr N_0$. The first index in each of these gives the number of pairs of through-lines, and the second index will be explained shortly. The modules $\mathscr W_{j,1}$ are called standard modules---they are modules over the Temperley--Lieb algebra when the multiplication rule is modified as follows: if the action of a basis diagram of the Temperley--Lieb algebra on a link state results in a link state with fewer through-lines, then the result is zero (this is precisely the meaning of the quotient). The standard modules $\mathscr W_{j,1}$ are irreducible representations. (As before, all modules ought to be additionally labeled with $N$ to avoid ambiguities, but we leave $N$ as implicit---this omission is common in the literature.)

\subsubsection{Affine Temperley--Lieb algebra}

With periodic boundary conditions, it is natural to introduce a translation $\tau$. In diagrammatic terms, using the labels of the right hand side of Eq.~\eqref{eq:e_j_diagram}, $\tau$ consists of the diagram where site $i$ of the bottom row is connected to site $i+1 \pmod{N}$ of the top row for $i = 1,\ldots, N$. The following additional defining relations are then obeyed diagrammatically:
\begin{subequations} \label{eq:TL_relations_affine}
\begin{gather}
\tau e_j \tau^{-1} = e_{j+1}, \qq{and} \label{eq:TL_translation_1} \\
\tau^2 e_{N-1} = e_1\cdots e_{N-1}.
\end{gather}
\end{subequations}
Moreover, $\tau^{\pm N}$ are central elements. The algebra generated by the generators $e_i$ and $\tau^{\pm 1}$, together with these relations in addition to \eqref{eq:TL_relations}, is usually called the affine Temperley--Lieb algebra $\ATL{N}(\m)$. The affine Temperley--Lieb algebra is infinite-dimensional---since there are no rules for noncontractible loops, an arbitrary number of noncontractible loops can be placed in a diagram with no through-lines, each giving rise to a distinct element of the algebra.

The standard modules $\mathscr W_{j,1}$ are also representations of $\ATL{N}(\m)$, and are irreducible for $j > 0$ and generic $\q$, though there are others. Importantly, $\mathscr W_{0,\q^{\pm 2}}$ is not irreducible---its submodule structure will be explained following the construction of the other standard modules. The other standard modules are obtained as follows. For a given value of $j > 0$, it is possible, using the action of the algebra, to cyclically permute the free sites on the top row of a basis diagram of $\ATL{N}(\m)$: this gives rise to the introduction of a pseudomomentum, which is parametrized by $\phi$. Whenever a diagram has $2j$ through-lines winding counterclockwise around the annulus $l$ times, this diagram may be replaced by one with the same connections but with the through-lines unwound, multiplied by a numerical factor $\e^{\i j l \phi}$. Stated more simply, there is a phase $\e^{\i\phi/2}$ for each complete counterclockwise winding of a through-line; however, it is impossible to unwind a single through-line on its own without colliding with other through-lines. Similarly, for $2j$ lines winding clockwise $l$ times, the numerical factor is $\e^{-\i j l \phi}$ when they are unwound, or $\e^{-\i\phi/2}$ per through-line per complete winding.

We will use an equivalent representation, where the phase is evenly graduated over the course of unwinding through-lines, instead of abruptly applying it to each complete winding. Instead of a phase $\e^{\pm\i\phi/2}$ when a through-line makes a complete winding (i.e., $N$ sites), we instead apply a phase $\e^{\pm\i\phi/2N}$ for each site a through-line moves right or left. This formulation preserves invariance under the translation operator. The rules regarding the phases associated to through-lines translate to the standard modules $\mathscr W_{j,\e^{\i\phi}}$ as follows. When a $\ATL{N}(\m)$ basis diagram is stacked on top of a link state, the phase $\e^{\pm\i\phi/2N}$ is now attributed for each step right or left of a through-line, as before. Finally, a weight must be assigned to noncontractible loops, which may differ from the weight for contractible loops. (The noncontractible loops are those already present in the affine Temperley--Lieb diagrams; for $j > 0$ the action of a diagram on a link state cannot generate additional noncontractible loops without eliminating the through-lines.)

The dimensions of these modules $\mathscr W_{j,\e^{\i\phi}}$ are then easily found by counting the link states. They are given by
\begin{equation} \label{eq:W_j_dimension}
\hat d_j = \binom{N}{N/2 + j}.
\end{equation}
Note that these dimensions do not depend on $\phi$, but representations with different $\e^{\i\phi}$ are not isomorphic.

The standard modules $\mathscr W_{j,\e^{\i\phi}}$ are irreducible for generic values of $\mathfrak q$ and $\phi$. However, degeneracies appear whenever the following \emph{resonance criterion} is satisfied \cite{MartinSaleur1993,GrahamLehrer1998}:
\begin{equation} \label{eq:resonance}
\exists k \in \mathbb N^* : \e^{\i\phi} = \mathfrak q^{2j+2k}.
\end{equation}
The representation $\mathscr W_{j,\mathfrak q^{2j+2k}}$ then becomes reducible, and contains a submodule isomorphic to $\mathscr W_{j+k,\mathfrak q^{2j}}$. The quotient $\mathscr W_{j,\mathfrak q^{2j+2k}}/\mathscr W_{j+k,\mathfrak q^{2j}}$ is generically irreducible, with dimension
\begin{equation} \label{eq:W_j_irreducible_dimension}
\overbar d_j \equiv \hat d_j - \hat d_{j+k}.
\end{equation}

There are also standard representations for $j = 0$---i.e., no through-lines. There is no pseudomomentum, but representations are still characterized by a second parameter, which now specifies the weight given to noncontractible loops generated by the action of the algebra on a link state, which are not possible for $j > 0$. Parametrizing this weight as $z + z^{-1}$, the corresponding standard module is denoted $\mathscr W_{0,z^2}$. This module is isomorphic to $\mathscr W_{0,z^{-2}}$, so we denote both henceforth by $\mathscr W_{0,z^{\pm 2}}$. If we identify $z = \e^{\i\phi/2}$, the resonance criterion of Eq.~\eqref{eq:resonance} still applies.

It is natural to require that $z + z^{-1} = \m$, so that contractible and noncontractible loops get the same weight. Imposing this requirement leads to the module $\mathscr W_{0,\mathfrak q^{\pm 2}}$, which is reducible even for generic $\mathfrak q$. Indeed, Eq.~\eqref{eq:resonance} is satisfied with $j = 0$ and $k = 1$, and hence $\mathscr W_{0,\mathfrak q^{\pm 2}}$ contains a submodule isomorphic to $\mathscr W_{1,1}$. Taking the quotient $\mathscr W_{0,\mathfrak q^{\pm 2}}/\mathscr W_{1,1}$ leads to a simple module for generic $\mathfrak q$, denoted by $\overbar{\mathscr W}_{\!\!0,\mathfrak q^{\pm 2}}$. It has dimension
\begin{equation}
\overbar d_0 = \hat d_0 - \hat d_1 = \binom{N}{N/2} - \binom{N}{N/2 + 1},
\end{equation}
in agreement with Eqs.~\eqref{eq:W_j_dimension} and \eqref{eq:W_j_irreducible_dimension}, so that we may extend the definitions of both to $j = 0$. An explicit description of $\overbar{\mathscr W}_{\!\!0,\mathfrak q^{\pm 2}}$ is given at the end of the next section.

\subsubsection{Jones--Temperley--Lieb algebras}

One may restrict to even powers of $\tau$. This is a natural choice from the point of view of the loop--cluster formulation since loops then appear as objects surrounding clusters, and thus only occur in even numbers, so that the loop model can be considered as defined on an oriented (often referred to as Chalker--Coddington) lattice \cite{ReadSaleur2007}. Formally, the elements $\tau^{\pm 2}$ are adjoined to the algebra instead of $\tau^{\pm 1}$, and Eq.~\eqref{eq:TL_translation_1} is replaced by
\begin{equation} \label{eq:TL_translation_2}
\tau^2 e_j \tau^{-2} = e_{j+2}.
\end{equation}
For this algebra, we give noncontractible loops the same weight $\m$ as for contractible loops. Finally, we take a quotient by the ideal generated by $\tau^N - 1$. In terms of annular diagrams, one annular diagram becomes equivalent to another obtained from the first by rotating the inner boundary through a full rotation, with the sites and attached curves following along with it. The resulting object with the modifications described in this paragraph is called the \emph{augmented Jones--Temperley--Lieb algebra} $JTL^{\text{au}}_N(\m)$, and it is finite-dimensional \cite{Gainutdinov2015}. $JTL^{\text{au}}_N(\m)$ is slightly larger than the \emph{Jones--Temperley--Lieb algebra} $JTL_N(\m)$, introduced by \textcite{ReadSaleur2007} and further studied by \textcite{GRS2013b}. The difference is entirely in the ideal with zero through-lines---i.e., the annular diagrams in which points of the outer boundary are paired together and points of the inner boundary are paired together (again, using curves that do not intersect), and no point of the inner boundary is paired with a point of the outer boundary. In $JTL^{\text{au}}_N(\m)$, all such diagrams are allowed. In $JTL_N(\m)$, only diagrams that can be drawn without crossing the periodic ``boundary'' between both pairs of sites $N$ and $1$ are allowed---in other words, the diagrams that can just as well be drawn on the (nonperiodic) rectangle. This ideal (the subalgebra of $JTL_N(\m)$ with zero through-lines) is also known as the oriented Jones annular subalgebra in the Brauer algebra \cite{Jones1994}. Formally, there is a covering homomorphism (surjection) of algebras $\psi: JTL^{\text{au}}_N(\m) \to JTL_N(\m)$, which acts nontrivially only in the zero through-lines subalgebra (of $JTL^{\text{au}}_N(\m)$). Its action is best understood through an example:
\begin{equation}\label{eq:psi_ex}
\psi: \quad
 \raisebox{-6mm}{\begin{tikzpicture}
 	\draw[thick, dotted] (-0.05,0.5) arc (0:10:0 and -7.5);
 	\draw[thick, dotted] (-0.05,0.55) -- (2.65,0.55);
 	\draw[thick, dotted] (2.65,0.5) arc (0:10:0 and -7.5);
	\draw[thick, dotted] (-0.05,-0.85) -- (2.65,-0.85);
	\draw[thick] (0,0.1) arc (-90:0:0.5 and 0.4);
	\draw[thick] (0,-0.1) arc (-90:0:0.9 and 0.6);
	\draw[thick] (2.6,-0.1) arc (-90:0:-0.9 and 0.6);
	\draw[thick] (2.6,0.1) arc (-90:0:-0.5 and 0.4);
	\draw[thick] (0.5,-0.8) arc (0:90:0.5 and 0.5);
	\draw[thick] (1.8,-0.8) arc (0:180:0.5 and 0.5);
	\draw[thick] (2.1,-0.8) arc (0:90:-0.5 and 0.5);
	\end{tikzpicture}}
	\quad\mapsto\quad
 \raisebox{-6mm}{\begin{tikzpicture}
 	\draw[thick, dotted] (-0.05,0.5) arc (0:10:0 and -7.5);
 	\draw[thick, dotted] (-0.05,0.55) -- (2.65,0.55);
 	\draw[thick, dotted] (2.65,0.5) arc (0:10:0 and -7.5);
	\draw[thick, dotted] (-0.05,-0.85) -- (2.65,-0.85);
	\draw[thick] (0.5,0.5) arc (-180:0:0.8 and 0.56);
	\draw[thick] (0.8,0.5) arc (-180:0:0.5 and 0.4);
	\draw[thick] (2.1,-0.8) arc (0:180:0.8 and 0.56);
	\draw[thick] (1.8,-0.8) arc (0:180:0.5 and 0.4);
	\end{tikzpicture}}
\end{equation}
The homomorphism $\psi$ essentially redraws (uniquely) the curves that keep the same pairs matched, but without crossing the boundary of the framing rectangle.

The difference between $JTL^{\text{au}}_N(\m)$ and $JTL_N(\m)$ is analogous to the difference between $\mathscr W_{0,\mathfrak q^{\pm 2}}$ and $\overbar{\mathscr W}_{\!\!0,\mathfrak q^{\pm 2}}$. Consider the case $N = 2$---i.e., the loop model for a two-site system, in the sector with no through-lines and with noncontractible loops given the same weight $\m = \mathfrak q + \mathfrak q^{-1}$ as contractible ones. Let us first write the two elements of the Temperley--Lieb algebra in the basis of the two link states $v_1 = (12)$ and $v_2 = (21)$:
\begin{subequations}
\begin{gather}
e_1 = \m \begin{pmatrix}
1 & 1 \\
0 & 0
\end{pmatrix}, \qq{and} \\
e_2 = \m \begin{pmatrix}
0 & 0 \\
1 & 1 \end{pmatrix}.
\end{gather}
\end{subequations}
Clearly, $e_1(v_1 - v_2) = e_2(v_1 - v_2) = 0$. Meanwhile, at $N = 2$ the action of $e_1$ and $e_2$ on the single state $(1)(2)$ in $\mathscr W_{1,1}$ is zero by definition, since the number of through-lines would decrease. Thus we see that $\mathscr W_{0,\mathfrak q^{\pm 2}}$ admits a submodule, generated by $v_1 - v_2$, that is isomorphic to $\mathscr W_{1,1}$. Diagrammatically, using what is technically called a \emph{Loewy diagram}, we have
\begin{equation} \label{eq:standard}
\mathscr W_{0,\mathfrak q^{\pm 2}}: \begin{tikzcd}
\overbar{\mathscr W}_{\!\!0,\mathfrak q^{\pm 2}} \arrow[d] \\
\mathscr W_{1,1} 
\end{tikzcd}.
\end{equation}
In such a diagram, the bottom module is a submodule, while the top module is a quotient module. The arrow indicates that within the standard module $\mathscr W_{0,\mathfrak q^{\pm 2}}$ a state in $\mathscr W_{1,1}$ can be reached from a state in $\overbar{\mathscr W}_{\!\!0,\mathfrak q^{\pm 2}}$ through the action of the Temperley--Lieb algebra, but the opposite is impossible.

For $N = 2$ sites, $\overbar{\mathscr W}_{\!\!0,\mathfrak q^{\pm 2}}$ is obtained via a quotient of $\mathscr W_{0,\mathfrak q^{\pm 2}}$ by $v_1 - v_2$. More colloquially, within $\overbar{\mathscr W}_{\!\!0,\mathfrak q^{\pm 2}}$, $(12)$ and $(21)$ are identified, so that any state in the zero through-lines sector can be rewritten in terms of diagrams that do not cross the periodic boundary. This feature extends to larger $N$: $\overbar{\mathscr W}_{\!\!0,\mathfrak q^{\pm 2}}$ consists of all link states with zero through-lines that do not cross the periodic boundary. Thus a basis of $\overbar{\mathscr W}_{\!\!0,\mathfrak q^{\pm 2}}$ for $N = 4$ is $\{(12)(34), (14)(23)\}$. Because this description also characterizes $JTL_N(\m)$, $\overbar{\mathscr W}_{\!\!0,\mathfrak q^{\pm 2}}$ is the natural building block for the natural representations of $JTL_N(\m)$, and we may use the same symbol $\psi$ to denote the homomorphism $\psi:\mathscr W_{0,\mathfrak q^{\pm 2}}\to\overbar{\mathscr W}_{\!\!0,\mathfrak q^{\pm 2}}$. We denote the natural representations of $JTL_N(\m)$ by $\overbar{\mathscr N}_{\!\!j} = \psi\mathscr N_j$, which we define as the collection of all link states with $2j$ or fewer through-lines, except that in the case of zero through-lines, only those link diagrams that do not cross the periodic boundary are included. Within these modules, whenever the action of the algebra produces a state with zero through-lines, it is redrawn with the same connections but without crossing the boundary (e.g., for $N = 2$, $e_2(1)(2) = (21) = (12)$). Note that this action is only well-defined because contractible and noncontractible loops are given the same weight. The collection of natural representations likewise admits a filtration:
\begin{equation}
\overbar{\mathscr N}_{\!\!N/2} \supset \overbar{\mathscr N}_{\!\!N/2 - 1} \supset \cdots \supset \overbar{\mathscr N}_{\!\!j} \supset \cdots \supset \overbar{\mathscr N}_{\!\!0} \supset 0.
\end{equation}
Now, the factors of the filtration, $\overbar{\mathscr N}_{\!\!j}/\overbar{\mathscr N}_{\!\!j-1} = \mathscr W_{j,1}$ and $\overbar{\mathscr N}_{\!\!0} = \overbar{\mathscr W}_{\!\!0,\mathfrak q^{\pm 2}}$, are irreducible for generic $\q$. We note that $\overbar{\mathscr N}_{\!\!2}$ will be especially important in what follows; $\overbar{\mathscr N}_{\!\!2} = \overbar{\mathscr W}_{\!\!0,\mathfrak q^{\pm 2}} + \mathscr W_{1,1} + \mathscr W_{2,1}$, where $+$ in contrast to $\oplus$ indicates that the right side is a vector space direct sum, but not a module direct sum. The modules $\overbar{\mathscr N}_{\!\!j}$ may also be referred to as ``glued modules'' as they are obtained from gluings of the standard modules.

\sectionbreak

The discussion of this section can be abstracted to a more general class of algebras whose basis elements can be viewed as the conjunction of two elements of a simpler set and whose multiplication follows certain rules analogous to the ones above. Such axioms define what are known as \emph{cellular algebras}, whose irreducible representations can be constructed systematically. The Temperley--Lieb algebra is the prototypical example of a cellular algebra.

\subsection{Identification of fields and operators on the lattice} \label{field_identification}

A key component of our approach is the control of the correspondence between objects on the lattice and in the CFT. 
In the following, we will rely on the so-called ``Koo--Saleur'' lattice construction of the Virasoro algebra generators. Recall that, following \cite{KooSaleur1994}, it is widely believed \cite{ZiniWang2018,GSJS2021} that 
\begin{gather}
\KSL_n = -\frac{L}{2\pi v_F}\sum_{j=1}^{2L}\e^{\i nj\pi/L}\left(e_j - e_\infty + \frac{\i}{v_F}[e_j, e_{j+1}]\right) + \frac{c}{24}\delta_{n,0}, \\
\KSLb_n = -\frac{L}{2\pi v_F}\sum_{j=1}^{2L}\e^{-\i nj\pi/L}\left(e_j - e_\infty - \frac{\i}{v_F}[e_j, e_{j+1}]\right) + \frac{c}{24}\delta_{n,0}\label{KSexpr}
\end{gather}
furnish representations of $\mathrm{Vir}$ and $\overbar{\mathrm{Vir}}$ in the limit $L\to\infty$ when restricted to scaling states. In the preceding equations, $e_j$ are the Temperley--Lieb generators described in Section \ref{sec:TL}, and 
\begin{equation}
e_\infty = \sin\gamma\int_{-\infty}^\infty\!\frac{\sinh[(\pi-\gamma)t]}{\sinh(\pi t)\cosh(\gamma t)}\,\mathrm dt
\end{equation}
is a constant shift that represents the average ground-state value of $e_j$. $v_F = \pi\sin\gamma/\gamma$ is the ``Fermi velocity'' or ``sound velocity'' whose value is known exactly from the Bethe-ansatz \cite{GSJS2021}. While the precise nature of the convergence is subtle and under active study, numerical and analytical checks have demonstrated that many of the expected properties of the Virasoro algebra are indeed recovered using (\ref{KSexpr}) ---see \cite{GSJS2021} for more detail. Thus, an eigenstate of $\KSL_0$ ($\KSLb_0$) with eigenvalue $h(L)$ ($\overbar h(L)$) can be viewed as the lattice representation on $2L$ sites of a field with left (right) conformal dimension $h\equiv\lim_{L\to\infty}h(L)$ ($\overbar h\equiv\lim_{L\to\infty}\overbar h(L)$).

In the following, we will use the notation
\begin{equation}
 H_n^{(L)}\equiv \KSL_n+\KSLb_{-n}.
\end{equation}
Most conformal identities involving primary fields and Virasoro generators other than $L_0$ and $\overbar L_0$ hold just as well when $L_{-n}$ is replaced by $L_{-n}+\overline{L}_n$ and $\overbar L_{-n}$ is replaced by $L_n+\overline{L}_{-n}$, since the Virasoro generator with positive index annihilates primary fields. Nevertheless, on the lattice, where this observation does not exactly hold, using this symmetric combination tends to improve convergence. The scaling dimensions $h(L) + \overbar h(L)$ are obtained by diagonalizing $H_0^{(L)} = \KSL_0 +\KSLb_0$. The precise identification of conformal fields is a subtle process because the convergence $h(L) \to h$ and $\overbar h(L) \to \overbar h$ can be slow, because these values are discretely parametrized by $L$ rather than a continuous parameter, making them more difficult to follow, and because the number of states proliferates rapidly with increasing $L$. To label the scaling states correctly, one must carefully follow sequences of eigenvalues $h(L)$ with increasing $L$, paying attention to other characteristics such as momentum and parity. The general methodology is explained in Appendix A.5 of \cite{JacobsenSaleur2019}. For our present purposes, we are for the most part interested in lattice analogues of fields within standard modules that are among the lowest excitations of the Hamiltonian in their respective sectors (equivalently, among the largest eigenvalues of the transfer matrix). Their identifications thus pose no unexpected challenges, and we typically label them with their positions within the spectrum. At the same time, we also reference their positions within Tables 13 and 14 of \cite{Grans-Samuelsson2020}, which carries out the field identification process in great detail for the particular (generic) case $Q = 1/2$---this shows where the fields fit in among others with the same $j$ in the modules $\mathscr W_{j,z^2}$.

\subsection{Scalar products}

\subsubsection{Loop scalar product: irreducible modules}

The loop scalar product $\langle\cdot,\cdot\rangle$ (or $\ip*{\cdot}$; we will systematically use this second notation below) is obtained by gluing the mirror image of one link state on top of the other and evaluating the result according to certain rules that we now describe. As usual, a factor $\m$ is given to each loop. However, unless all through-lines connect through from bottom to top the result is zero. We also take into account the weight of straightening the connected through-lines: a through-line that has moved to the right (left) is assigned the weight $\e^{\i \phi/2N}$ ($\e^{-\i \phi/2N}$) for each step, where $\phi$ is again the same parameter that appears in $\mathscr W_{j,\e^{\i\phi}}$. Importantly, the ex ante inner product of any two states is nonzero---if the diagram produced by gluing two link states produces no loops and movement of through-lines, its inner product is $\m^0 (\e^{\i \phi/2N})^0 = 1$. To illustrate this scalar product we take the following examples, where the solid lines around the rightmost diagrams signify that we assign them a value according to the aforementioned rules:
\begin{subequations} \label{eq:ip_examples}
\begin{gather}
\braket*{(12)(3)(4)}{(1)(2)(34)} = \left\langle\;\vcenter{\hbox{\begin{tikzpicture}
\newcommand{\dist}{0.2}
\draw[thick] (0,0) arc (-180:0:\dist);
\draw[thick] (4*\dist,0) -- (4*\dist,-2*\dist);
\draw[thick] (6*\dist,0) -- (6*\dist,-2*\dist);
\draw[thick, dotted] ($(current bounding box.north east) + (0.05+\dist,0.05)$) rectangle ($(current bounding box.south west)+ (-0.05-\dist,-0.05)$);
\end{tikzpicture}}}
\;\middle|\;
\vcenter{\hbox{\begin{tikzpicture}
\newcommand{\dist}{0.2}
\draw[thick] (4*\dist,0) arc (-180:0:\dist);
\draw[thick] (2*\dist,0) -- (2*\dist,-2*\dist);
\draw[thick] (0,0) -- (0,-2*\dist);
\draw[thick, dotted] ($(current bounding box.north east) + (0.05+\dist,0.05)$) rectangle ($(current bounding box.south west)+ (-0.05-\dist,-0.05)$);
\end{tikzpicture}}}
\;\right\rangle
=\;
\vcenter{\hbox{\begin{tikzpicture}
\newcommand{\dist}{0.2}
\begin{scope}[yscale=-1,xscale=1]
\draw[thick] (0,0) arc (-180:0:\dist);
\draw[thick] (4*\dist,0) -- (4*\dist,-2*\dist);
\draw[thick] (6*\dist,0) -- (6*\dist,-2*\dist);
\end{scope}
\draw[thick] (4*\dist,0) arc (-180:0:\dist);
\draw[thick] (2*\dist,0) -- (2*\dist,-2*\dist);
\draw[thick] (0,0) -- (0,-2*\dist);
\draw ($(current bounding box.north east) + (0.05+\dist,0.05)$) rectangle ($(current bounding box.south west)+ (-0.05-\dist,-0.05)$);
\end{tikzpicture}}}
\; = 0 , \label{ip_example_1} \\
\begin{aligned}
\braket*{(14)(23)(5)(6)}{(1)(23)(45)(6)} &= \left\langle\;\vcenter{\hbox{\begin{tikzpicture}
\newcommand{\dist}{0.2}
\draw[thick] (2*\dist,0) arc (-180:0:\dist);
\draw[thick] (0,0) arc (-180:0:3*\dist);
\draw[thick] (8*\dist,0) -- (8*\dist,-3*\dist);
\draw[thick] (10*\dist,0) -- (10*\dist,-3*\dist);
\draw[thick, dotted] ($(current bounding box.north east) + (0.05+\dist,0.05)$) rectangle ($(current bounding box.south west)+ (-0.05-\dist,-0.05)$);
\end{tikzpicture}}}
\;\middle|\;
\vcenter{\hbox{\begin{tikzpicture}
\newcommand{\dist}{0.2}
\draw[thick] (2*\dist,0) arc (-180:0:\dist);
\draw[thick] (6*\dist,0) arc (-180:0:\dist);
\draw[thick] (0,0) -- (0,-2*\dist);
\draw[thick] (10*\dist,0) -- (10*\dist,-2*\dist);
\draw[thick, dotted] ($(current bounding box.north east) + (0.05+\dist,0.05)$) rectangle ($(current bounding box.south west)+ (-0.05-\dist,-0.05)$);
\end{tikzpicture}}}
\;\right\rangle \\
&= \;\vcenter{\hbox{\begin{tikzpicture}
\newcommand{\dist}{0.2}
\begin{scope}[yscale=-1,xscale=1]
\draw[thick] (2*\dist,0) arc (-180:0:\dist);
\draw[thick] (0,0) arc (-180:0:3*\dist);
\draw[thick] (8*\dist,0) -- (8*\dist,-3*\dist);
\draw[thick] (10*\dist,0) -- (10*\dist,-3*\dist);
\end{scope}
\draw[thick] (2*\dist,0) arc (-180:0:\dist);
\draw[thick] (6*\dist,0) arc (-180:0:\dist);
\draw[thick] (0,0) -- (0,-2*\dist);
\draw[thick] (10*\dist,0) -- (10*\dist,-2*\dist);
\draw ($(current bounding box.north east) + (0.05+\dist,0.05)$) rectangle ($(current bounding box.south west)+ (-0.05-\dist,-0.05)$);
\end{tikzpicture}}}
\;= \e^{2\i \phi/N} \m = \e^{\i \phi/3} \m,
\end{aligned} \label{ip_example_2} \\
\braket*{(23)(41)}{(14)(23)} = \left\langle\;\vcenter{\hbox{\begin{tikzpicture}
\newcommand{\dist}{0.2}
\draw[thick] (\dist,-\dist) arc (-90:0:\dist);
\draw[thick] (4*\dist,0) arc (-180:0:\dist);
\draw[thick] (9*\dist,-\dist) arc (-90:-180:\dist);
\draw[thick, dotted] ($(current bounding box.north east) + (0.05,0.05)$) rectangle ($(current bounding box.south west)+ (-0.05,-0.05-1*\dist)$);
\end{tikzpicture}}}
\;\middle|\;
\vcenter{\hbox{\begin{tikzpicture}
\newcommand{\dist}{0.2}
\draw[thick] (2*\dist,0) arc (-180:0:\dist);
\draw[thick] (0,0) arc (-180:0:3*\dist);
\draw[thick, dotted] ($(current bounding box.north east) + (0.05+\dist,0.05)$) rectangle ($(current bounding box.south west)+ (-0.05-\dist,-0.05)$);
\end{tikzpicture}}}
\;\right\rangle
= \;
\vcenter{\hbox{\begin{tikzpicture}
\newcommand{\dist}{0.2}
\begin{scope}[xshift=2*\dist cm]
\draw[thick] (2*\dist,0) arc (-180:0:\dist);
\draw[thick] (0,0) arc (-180:0:3*\dist);
\end{scope}
\begin{scope}[yscale=-1,xscale=1]
\draw[thick] (\dist,-\dist) arc (-90:0:\dist);
\draw[thick] (4*\dist,0) arc (-180:0:\dist);
\draw[thick] (9*\dist,-\dist) arc (-90:-180:\dist);
\end{scope}
\draw ($(current bounding box.north east) + (0.05,0.05)$) rectangle ($(current bounding box.south west)+ (-0.05,-0.05)$);
\end{tikzpicture}}}
= \m^2. \label{ip_example_3}
\end{gather}
\end{subequations}

This loop scalar product is then extended by sesquilinearity to the whole space of loop states. The adjoint $U^\dagger$ of a word $U$ in the Temperley--Lieb algebra can be defined similarly by flipping the diagram representing it about a horizontal line, as in the following example:
\begin{equation}
\vcenter{\hbox{\begin{tikzpicture}
 \newcommand{\dist}{0.2}
 	\draw[thick] (2*\dist,0) arc (-180:0:\dist);
	\draw[thick] (0,0) arc (-180:0:3*\dist);
	\draw[thick] (8*\dist,0) -- (8*\dist,-3*\dist);
	\draw[thick] (10*\dist,0) -- (10*\dist,-3*\dist);
 \begin{scope}[xshift=0cm,yshift=-5*\dist cm]
 \begin{scope}[yscale=-1,xscale=1]
 	\draw[thick] (4*\dist,0) arc (-180:0:\dist);
	\draw[thick] (0,0) arc (-180:0:\dist);
	\draw[thick] (8*\dist,0) -- (8*\dist,-3*\dist);
	\draw[thick] (10*\dist,0) -- (10*\dist,-3*\dist);
 \end{scope}
 \end{scope}
\draw[thick,dotted] ($(current bounding box.north east) + (0.05+\dist,0.05)$) rectangle ($(current bounding box.south west)+ (-0.05-\dist,-0.05)$);
\end{tikzpicture}}} 	
	^\dagger
	= \;\;
	\vcenter{\hbox{\begin{tikzpicture}
 \newcommand{\dist}{0.2}
 \begin{scope}[yscale=-1,xscale=1]
 	\draw[thick] (2*\dist,0) arc (-180:0:\dist);
	\draw[thick] (0,0) arc (-180:0:3*\dist);
	\draw[thick] (8*\dist,0) -- (8*\dist,-3*\dist);
	\draw[thick] (10*\dist,0) -- (10*\dist,-3*\dist);
 \begin{scope}[xshift=0cm,yshift=-5*\dist cm]
 \begin{scope}[yscale=-1,xscale=1]
 	\draw[thick] (4*\dist,0) arc (-180:0:\dist);
	\draw[thick] (0,0) arc (-180:0:\dist);
	\draw[thick] (8*\dist,0) -- (8*\dist,-3*\dist);
	\draw[thick] (10*\dist,0) -- (10*\dist,-3*\dist);
 \end{scope}
 \end{scope}
 \end{scope}
\draw[thick,dotted] ($(current bounding box.north east) + (0.05+\dist,0.05)$) rectangle ($(current bounding box.south west)+ (-0.05-\dist,-0.05)$);
 \end{tikzpicture}}}.
\end{equation}
From this definition it is clear that the generators $e_i$ themselves are self-adjoint. By construction, this loop scalar product is invariant with respect to the Temperley--Lieb action: $\ip*{x}{Uy}=\ip*{U^\dagger x}{y}$, where $U$ is a word in the Jones--Temperley--Lieb algebra. The loop scalar product is of course not positive definite. It is however not degenerate (provided $\m\ne 0$). 

The loop scalar product is the simple translation of the natural $U(\m)$ scalar product \cite{ReadSaleur2007} to the geometrical setting. It is also the natural scalar product appearing in lattice correlation functions, and as such of course it goes over to the conformal scalar product in the continuum limit \cite{DJS2010c}. Recall that this scalar product obeys in particular $L_n^\dagger=L_{-n}$. For these reasons we will also use the terminology ``conformal scalar product'' when referring to the loop scalar product on the lattice.

\subsubsection{Loop scalar product: natural or glued modules}

Within the natural representations we define an inner product that is mostly the same as in the natural case, though we drop the restriction that all through-lines must connect through from bottom to top for a nonzero result. Within $\mathscr N_j$ with $j \ge 1$, for instance, the example \eqref{ip_example_1} would have a value of $1$ instead of $0$. Since these natural modules have no pseudomomentum, there is no phase attributed to the lateral movement of through-lines: $\phi = 0$. The inner product in the example of \eqref{ip_example_2} is thus $\e^{\i\phi/3}\m = \m$, and \eqref{ip_example_3} remains unchanged. The JTL generators $e_j$ remain self-adjoint with respect to this inner product.

\subsubsection{Positive-definite scalar product}\label{subsec:posdefscalprod}

Since the loop scalar product is not positive definite, it will be crucial in what follows to introduce another scalar product that has these properties. We choose here the ``native'' product for which basic loop states are mutually orthogonal and of unit norm square, and denote this product by $(\cdot,\cdot)$ (or $(\cdot|\cdot)$; we will systematically use this second notation outside of this section, to avoid confusion when writing vector components). This scalar product coincides with the formal limit $\m\to\infty$ of the loop scalar product, after proper renormalization in each sector with a given number of through-lines. We will also refer to this scalar product as the Euclidean scalar product, and the induced norm as the Euclidean norm or 2-norm. Note that we do not discuss here the meaning of this scalar product in the conformal field theory itself. 

\subsection{Continuum limits: generating functions}

While the results below appear in several of our papers already, they are useful to set up the context. We begin with the standard parametrization $x\in (0,\infty]$, in terms of which the loop weight is
\begin{equation}
 \m = \sqrt Q = 2\cos\left(\frac{\pi}{x+1}\right),
\end{equation}
and the corresponding central charge 
\begin{equation}
 c = 1 - \frac{6}{x(x+1)},
\end{equation}
We will parametrize exponents using the Kac formula 
\begin{equation}
 h_{r,s} = \frac{[r(x+1) - sx]^2 - 1}{4x(x+1)}.
\end{equation}
(although in many cases $r,s$ will not be integer).
Define
\begin{equation} \label{eq:F_characters}
F_{j,\e^{\i\phi}} = \frac{q^{-c/24}\overbar q^{-c/24}}{P(q)P(\overbar q)} \sum_{e\in\mathbb Z} q^{h_{e-e_\phi,-j}}\overbar q^{h_{e-e_\phi,j}},
\end{equation}
in which
\begin{equation}
P(q) = \prod_{n=1}^\infty (1-q^n)
\end{equation}
and $e_\phi = \phi/2\pi$. As usual, $q$ and $\overbar q$ are the modular parameters of the torus. $P(q)$, the inverse generating function for the partition numbers, can be written in terms of the Dedekind eta function as $P(q) = q^{-1/24}\eta(q)$.

The continuum-limit partition function for the $Q$-state Potts model is then given by
\begin{equation} \label{eq:torus_Z}
Z_Q = F_{0,\mathfrak q^{\pm 2}} + \frac{Q-1}{2}F_{0,-1} + \sum_{j>0} D_{j,0}F_{j,1} + \sum_{\substack{j>0,k>1\\k|j}} \sum_{\substack{0<p<k \\ p\wedge k = 1}} D_{j,\pi p/k} F_{j,\e^{2\pi\i p/k}},
\end{equation}
where $p\wedge k$ is the greatest common divisor of $p$ and $k$. The coefficients $D_{j,K}$ can be thought of as ``multiplicities,'' although for generic $Q$, they are not integers. Their interpretation in terms of symmetries is beyond the scope of this discussion \cite{Gainutdinov2015,JRS2022}. They are given by
\begin{equation}
D_{j,K} = \frac{1}{j}\sum_{r=0}^{j-1} \e^{2\i Kr} w(j,j\wedge r),
\end{equation}
($j\wedge 0 = j$ by definition) and
\begin{equation}
w(j,d) = \mathfrak q^{2d} + \mathfrak q^{-2d} + \frac{Q-1}{2}(\i^{2d} + \i^{-2d}) = \mathfrak q^{2d} + \mathfrak q^{-2d} + (-1)^d(Q-1).
\end{equation}

Equations \eqref{eq:F_characters} and \eqref{eq:torus_Z} encode the operator content of the $Q$-state Potts model CFT. The conformal weights arising from the last term in Eq.~\eqref{eq:torus_Z} are of the form
\begin{equation}
(h, \overbar h) = (h_{e-p/k,j},h_{e-p/k,-j}). \qquad (e\in\mathbb Z)
\end{equation}
The first two terms must be handled slightly differently. Using the identity
\begin{equation}
F_{0,\mathfrak q^{\pm 2}} - F_{1,1} = \sum_{n=1}^\infty K_{n,1}\overbar K_{n,1} \equiv \overbar F_{0,\mathfrak q^{\pm 2}}
\end{equation}
with the Kac character
\begin{equation}
K_{r,s} = q^{h_{r,s}-c/24}\frac{1-q^{rs}}{P(q)},
\end{equation}
we see that we get the set of diagonal fields
$(h_{n,1},h_{n,1})$ ($n\in\mathbb N^*$). The partition function can then be rewritten as
\begin{equation} \label{eq:torus_Z1}
Z_Q = \overbar F_{0,\mathfrak q^{\pm 2}} + \frac{Q-1}{2}F_{0,-1} + F_{1,1} + \sum_{j>0} D_{j,0}F_{j,1} + \sum_{\substack{j>0,k>1\\k|j}} \sum_{\substack{0<p<k \\ p\wedge k = 1}} D_{j,\pi p/k} F_{j,\e^{2\pi\i p/k}}.
\end{equation}

We notice now that $D_{1,0} = \mathfrak q^2 + \mathfrak q^{-2} - (Q-1) = Q - 2 - (Q - 1) = -1$. Hence $F_{1,1}$ disappears, in fact, from the partition function. $F_{1,1}$ corresponds geometrically to the so-called hull operator \cite{SaleurDuplantier1987}---related to the indicator function that a point is at the boundary of an FK cluster---with corresponding conformal weights $(h_{0,1},h_{0,1})$. Therefore, this operator is absent from the partition function. We will, nevertheless, continue to consider $\mathscr W_{1,1}$, since this module does appear in related models, such as the oriented loop (or $U(\m)$) model \cite{ReadSaleur2001}. We note meanwhile that the higher hull operators---related to the indicator function that $j > 1$ distinct hulls come close together at the scale of the lattice spacing---with conformal weights $(h_{0,j},h_{0,j})$ in $F_{j,1}$ do appear in the partition function.

The decomposition of the Potts-model partition function in Eq.~\eqref{eq:torus_Z1} for generic $Q$ is, in fact, in one-to-one correspondence with an algebraic decomposition---exact in finite size---of the Hilbert space $\mathscr H_Q$ in terms of modules of the affine Temperley--Lieb algebra $\ATL{N}(\m)$ \cite{ReadSaleur2007}. This decomposition formally reads
\begin{equation} \label{eq:torus_H}
\mathscr H_Q = \overbar{\mathscr W}_{\!\!0,\mathfrak q^{\pm 2}} + \frac{Q-1}{2}{\mathscr W}_{0,-1} + {\mathscr W}_{1,1} + \sum_{j>0} D_{j,0}{\mathscr W}_{j,1} + \sum_{\substack{j>0,k>1\\k|j}} \sum_{\substack{0<p<k \\ p\wedge k = 1}} D_{j,\pi p/k} {\mathscr W}_{j,\e^{2\pi\i p/k}}.
\end{equation}
Eq.~\eqref{eq:torus_H} is only formal in the sense that, for generic $Q$, the multiplicities are not integers, and $\mathscr H_Q$ cannot be interpreted as a proper vector space. By contrast, the modules $\mathscr W_{j,\e^{\i\phi}}$ are well-defined spaces with integer dimension independent of $Q$. Note that one must take into account the fact that the sums e.g. in (\ref{eq:torus_H}) must be properly truncated for a finite lattice system.

The torus partition function in Eq.~\eqref{eq:torus_Z} is obtained by a trace over $\mathscr H_Q$,
\begin{equation}
Z_Q = \tr_{\mathscr H_Q} \e^{-\beta_R H} \e^{-\i\beta_I P},
\end{equation}
where the real parameters $\beta_R > 0$ and $\beta_I$ determine the size of the torus, while $H$ and $P$ denote respectively the lattice Hamiltonian and momentum operators. Introducing the (modular) parameters
\begin{subequations}
\begin{gather}
q = \exp\left[-\frac{2\pi}{N}(\beta_R + \i\beta_I)\right] \qq{and} \\
\overbar q = \exp\left[-\frac{2\pi}{N}(\beta_R - \i\beta_I)\right],
\end{gather}
\end{subequations}
we have, in the limit where the size of the system $N\to\infty$, with $\beta_R,\beta_I\to\infty$ such that $q$ and $\overbar q$ remain finite,
\begin{equation}
\tr_{\mathscr W_{j,\e^{\i\phi}}} \e^{-\beta_R H} \e^{-\i\beta_I P} \overset{N\to\infty}{\mapsto} F_{j,\e^{\i\phi}},
\end{equation}
while in the $j = 0$ case the relevant continuum limit is
\begin{equation}
 \tr_{\overbar{\mathscr W}_{\!\!0,\q^{\pm 2}}}\e^{-\beta_R H} \e^{-\i\beta_I P} \overset{N\to\infty}{\mapsto} \overbar F_{0,\q^{\pm 2}},
\end{equation}
hence ensuring the correspondence between (\ref{eq:torus_H}) and (\ref{eq:torus_Z1}).

\subsection{Conjectured Virasoro modules} \label{conjectures}

In \cite{GSJLS} we proposed the following conjecture for generic $\q$:
\begin{conjecture}[Quotient loop-model module without through-lines] \label{loop_0}
In the scaling limit
\begin{equation}
\overbar{\mathscr W}_{\!\!0,\mathfrak q^{\pm 2}} \mapsto \bigoplus_{n=1}^\infty \IrrV{n,1}\otimes \IrrVb{n,1}.
\end{equation}
\end{conjecture}
We use the notation $\IrrV{r,s}$ to generally denote irreducible Virasoro modules with highest weight $h_{r,s}$. In the present case, these modules are simply Kac modules, with the submodule generated by the unique singular vector at level $rs$ factored out. This result does not involve Jordan blocks, and was well checked in \cite{Grans-Samuelsson2020}. We also proposed 
\begin{conjecture}[Loop-model modules with through-lines] \label{loop_j}
For $j > 0$ and $2j$ through lines we have the scaling limit
\begin{equation}
\mathscr W_{j,1} \mapsto (\Verma{0,-j} \otimes \Vermab{0,j}) \oplus \bigoplus_{e=1}^\infty \mathscr L_{e,j}.
\end{equation}
\end{conjecture}
Here, $\Verma{r,s}$ denotes the Verma module for highest weight $h_{r,s}$ with $r \notin \mathbb{N}^*$ or $s \notin \mathbb{N}^*$. As for the ${\mathscr L}_{e,j}$ they are more complicated objects whose structure was conjectured based mostly on bootstrap consistency arguments as follows. Considerations of the observed multiplicities of fields in the generating function and self-duality give rise to the following quartet of fields, the simplest structure consistent with both considerations:
\begin{equation} \label{eq:indecomposable_diamond}
\begin{tikzcd}
& \tilde\Psi_{e,j} \arrow[dr,"\overbar A^\dagger_{e,j}"]\arrow[dl,"A^\dagger_{e,j}"']\arrow[to=3-2,"L_0 - h_{e,-j}"] & \\
\Phi_{e,j} = \phi_{e,j}\otimes\overbar\phi_{e,-j} \arrow[dr,"A_{e,j}"'] & & \overbar\Phi_{e,j} = \phi_{e,-j}\otimes\overbar\phi_{e,j} \arrow[dl,"\overbar A_{e,j}"] \\
& \Psi_{e,j} \equiv A_{e,j}\Phi_{e,j} = \overbar A_{e,j}\overbar\Phi_{e,j}
\end{tikzcd}
\end{equation}
where $A_{e,j}$ is the operator in Virasoro theory creating a null state when acting on the highest weight state with conformal weight $h_{e,j}$. In this diagram, $\phi_{e,\pm j}$ and $\overbar{\phi}_{e,\pm j}$ are $\Vir$ and $\overbar{\Vir}$ primaries from which are formed the fields $\Phi_{e,j}$ and $\overbar{\Phi}_{e,j}$ of identical physical dimension $x_{e,j}=h_{e,j}+h_{e,-j}$. The fields $\tilde{\Psi}_{e,j}$ and $\Psi_{e,j} = A_{e,j}\Phi_{e,j}=\overbar{A}_{e,j}\overbar{\Phi}_{e,j}$ both have dimension $x_{e,j}+ej$. They are in rank two Jordan blocks for $L_0$ and $\overbar{L}_0$. The corresponding structure of Virasoro modules defines the quotient modules $\mathscr L_{e,j}$:
\begin{equation}
\begin{tikzcd}[column sep=small]
& & \Verma{e,-j}\otimes\Vermab{e,-j} \arrow[dr]\arrow[dl] & \\
\mathscr L_{e,j} \equiv \mathscr Q[(\Verma{e,j}^{(\text d)} \otimes\Vermab{e,-j})\oplus (\Verma{e,-j}\otimes\Vermab{e,j}^{(\text d)})]: & \IrrV{e,j}\otimes\Vermab{e,-j} \arrow[dr] & & \Verma{e,-j}\otimes\IrrVb{e,j} \arrow[dl] \\
& & \Verma{e,-j}\otimes\Vermab{e,-j}
\end{tikzcd}
\end{equation}
This gives
\begin{equation}
\begin{aligned}
\tr_{{\mathscr L }_{e,j}}q^{L_0}\overbar{q}^{\overbar{L}_0}&=2\frac{q^{h_{e,-j}}}{P(q)}\frac{\overbar{q}^{h_{e,-j}}}{P(\overbar{q})}+\frac{q^{h_{e,j}}-q^{h_{e,-j}}}{P(q)}\frac{\overbar{q}^{h_{e,-j}}}{P(\overbar{q})}+\frac{q^{h_{e,-j}}}{P(q)}\frac{\overbar{q}^{h_{e,j}}-\overbar{q}^{h_{e,-j}}}{P(\overbar{q})} \\
&=\frac{q^{h_{e,j}}}{P(q)}\frac{\overbar{q}^{h_{e,-j}}}{P(\overbar{q})}+\frac{q^{h_{e,-j}}}{P(q)}\frac{\overbar{q}^{h_{e,j}}}{P(\overbar{q})}.
\end{aligned}
\end{equation}

In \cite{Grans-Samuelsson2020} we checked numerically part of this picture---essentially, the bottom component of the structure (that is, the fact that $A_{e,j}\Phi_{e,j}$ and $\overbar A_{e,j}\overbar\Phi_{e,j}$ could be identified with a single field denoted $\Psi_{e,j}$). A crucial fact that remains to be confirmed is the existence of a Jordan block for $L_0$ and $\overbar{L}_0$. Such Jordan cells have only be observed so far in the special case of $\q$ a root of unity, where the representation theory of the $\JTL{2L}(\m)$ algebra becomes non-semisimple, so the lattice Hamiltonian itself exhibits Jordan blocks. The situation here is different since the Jordan cell is expected to only appear in the continuum limit. As we shall see, this situation can still be studied analytically and numerically. 

To each such diamond structure \eqref{eq:indecomposable_diamond} is associated an indecomposability parameter, $b_{e,j}$, such that
\begin{subequations}
\begin{gather}
(L_0 - h_{e,-j})\tilde\Psi_{e,j} = (\overbar L_0 - h_{e,-j})\tilde\Psi_{e,j} = A_{e,j}\Phi_{e,j} = \overbar A_{e,j}\overbar{\Phi}_{e,j}, \\
A_{e,j}A^\dagger_{e,j}\tilde\Psi_{e,j} = b_{e,j}\Psi_{e,j}, \qq{and} \\
\overbar A_{e,j}\overbar A^\dagger_{e,j} \tilde\Psi_{e,j} = b_{e,j}\overbar\Psi_{e,j}.
\end{gather}
\end{subequations}
They may be calculated as quotients, $b_{e,j} = \nu_{e,j}/\kappa_{e,j}$, where
\begin{subequations}
 \begin{gather}
 \nu_{e,j} = \lim_{\epsilon\to 0}\frac{\ev*{A^\dagger_{e,j}A_{e,j}}{\phi_{e+\epsilon,j}}}{\epsilon} \qq{and} \\
 \kappa_{e,j} = \lim_{\epsilon\to 0}\frac{h_{e+\epsilon,j} + ej - h_{e-\epsilon,-j}}{\epsilon} = \frac{e(x+1)}{x}.
 \end{gather}
\end{subequations}
For the special cases $(e,j) = (1,1)$ and $(1,2)$, routine calculation \cite{Grans-Samuelsson2020,NR_2021} gives
\begin{subequations}
 \begin{gather}
 b_{1,1} = \frac{1}{x+1} \qq{and} \label{eq:b11} \\
 b_{1,2} = \frac{4}{x+1} - \frac{2}{x^2}. \label{eq:b12}
 \end{gather}
\end{subequations}
Much of the remainder of this paper will be devoted to observing these Jordan blocks and measuring these $b$ values directly on the lattice model.

\section{The study of $L_0$ and $\overbar L_0$ Jordan blocks and the measurement of $b$} \label{strategy}

\subsection{Identification of emerging Jordan blocks}

The general issue we are facing is how to identify the formation of a Jordan block in the limit $L\to\infty$ in a set-up where, presumably, there is never such a block for $L$ finite. We are not aware of any theory about scenarios for the emergence of Jordan blocks as limits of diagonalizable matrices. In the case of blocks of rank two at least, it is easy to build such a scenario. Consider the matrix 
\begin{equation}
M(x) = \begin{pmatrix}
0 & 1 \\
0 & x \end{pmatrix}
\label{test_2}
\end{equation}
which is generically diagonalizable with eigenvalues $0$ and $x$ and (unnormalized) eigenvectors $u=(1,0)$ and $v=(1,x)$. As $x\to 0$, these two eigenvectors converge towards each other, while $M(x=0)$ is then a rank two Jordan block. 

To see whether two eigenvectors approach each other as $L\to\infty$, we must use a scalar product that is positive definite. Any such product will do of course, and in our case we will use the native loop scalar product defined earlier. Once we have identified (more on this below) a pair of eigenvalues and hence a rank two sub-matrix in the Hamiltonian $H$ for which we expect a Jordan block will appear in the continuum limit, we can, calling the corresponding eigenvectors $\{u, v\}$, consider the quotient
\begin{equation}
J(u,v) = \frac{(u|v)}{\sqrt{(u|u)(v|v)}}.
\end{equation}
By the Cauchy--Schwarz inequality, $|(u|v)|^2 \le (u|u)(v|v)$ implies $|J(u,v)| \le 1$ (equality holds if and only if the two vectors are parallel).

Let us discuss this on the simple example of the test matrix (\ref{test_2}). If $u$ and $v$ are the two eigenvectors $(1,0)$ and $(1,x)$ we find 
\begin{equation}
J(u,v) = \frac{(u|v)}{\sqrt{(u|u)(v|v)}} = \frac{1}{\sqrt{1+x^2}},
\end{equation}
which is clearly less than 1 for $x \ne 0$ and $J\to 1$ when $x \to 0$. 

Let us now discuss the generalized eigenvector at $x = 0$, which can be anything of the form $(y,1)$---we can choose $(0,1)$ so that it is orthogonal to $u=(1,0)$. If we take the pair $(u,v)$ at generic $x$ and perform Gram--Schmidt orthogonalization, we obtain $\hat u = u = (1,0)$ and $v'=(0,1)$, which is exactly what we are looking for---a basis of the Jordan form of the limiting matrix $M(0)$. We may formally denote this process using the following notation:
\begin{equation}
\mathrm{GS}[u,v] = (\hat u,v') \equiv ((1,0),(0,1))
\end{equation}
[the dependence of $v$ on $x$ has been suppressed]. Of course, for generic $x$, $v'$ is no longer an eigenvector.

If we were to apply the Gram--Schmidt procedure to $(v,u)$ instead we would find
\begin{equation}
\mathrm{GS}[v,u] = (\hat v,u') \equiv \Big(\Big(\frac{1}{\sqrt{x^2+1}},\frac{x}{\sqrt{x^2+1}}\Big),\Big(\frac{|x|}{\sqrt{x^2+1}},-\frac{\sgn x}{\sqrt{x^2+1}}\Big)\Big),
\end{equation}
and then
\begin{equation}
\lim_{x\to 0}(\hat v,u') = ((1,0),(0,\mp 1))
\end{equation}
(depending on whether the limit is approached from above or below), recovering again the true and generalized eigenvector.

The observations above that a pair of eigenvectors have a $J$ measure that converges to 1 and that the Gram--Schmidt procedure yields the generalized eigenvector are not specific to our carefully chosen example. They are general features of nondiagonalizable operators that are obtained as limits of sequences of diagonalizable operators, and that these observations indeed signal nondiagonalizability of the limit \cite{Liu2024}.

Thus, whenever a pair of vectors $\{u,v\}$ satisfies $J(u,v)\to 1$ in some limit, we will call the second vector obtained from performing the Gram--Schmidt procedure an emerging Jordan vector. The submatrix of the Hamiltonian in the basis of the orthogonalized vectors will be called an emerging Jordan block.

For $c$ generic, the rank two Jordan blocks that are expected in the continuum limit appear, on the lattice, as rank two diagonalizable matrices with slightly different eigenvalues. The critical exponents we can extract from these eigenvalues also differ slightly in finite size. As $L\to\infty$ these exponents converge to each other, and, if a rank two Jordan block is to appear in this limit, so should the eigenvectors---just like in the toy example above. In Appendix \ref{sec:bTt} we consider the $c \to 0$ limit for various lattice sizes $L$ and use the technique described to compute properties of the well-studied Jordan block that arises between the stress--energy tensor $T$ and its logarithmic partner $t$.

It should be emphasized that an isolated measurement of $J$, even if the value is near 1, is not particularly meaningful \emph{per se}. Nor will a $J$ value of 1 ever be observed, at least between two distinct eigenvectors, since a measurement of 1 contradicts the linear independence of distinct eigenvectors. What is, in fact, meaningful is the observation of a limiting value of 1. In effect, we are looking to infer the existence of a nontrivial Jordan block without resorting to a computation of the basis giving rise to the Jordan form itself. In principle, it is analytically possible to determine that there is a Jordan block by showing that the minimal polynomial has a repeated root, although this quickly becomes intractable with increasing dimension. Furthermore, numerical determination of the Jordan canonical form is notoriously unstable, and cannot be applied to our limiting situation where the underlying vector space is infinite-dimensional. We will see shortly that the emerging Jordan block technique can be applied numerically as well, bypassing all of these problems.

The fact that the continuum limit Jordan blocks sometimes already appear in finite size and some times not is not fully understood. Corrections to the continuum limit Hamiltonian due to irrelevant operators \cite{LUKYANOV2003323} (such as the square of the stress-energy tensor $:T^2:$) are enough in general to split $L_0$ Jordan blocks by shifting in an $L$ dependent way the two coinciding (pseudo)eigenvalues. But this is not always the case, since some of the (quantum group) symmetries of the CFT are already present in the lattice model. How to fully reconcile symmetries and corrections to scaling remains an open problem. 

\subsection{The measurement of $b$} \label{b_emerging}

Although this paper (except the appendix) is devoted to the generic case, let us, to fix ideas, first recall what happens at $c=0$ \cite{GurarieLudwig2002}. The indecomposability in the continuum limit manifests as non-diagonalizability of $L_0$ and $\overbar L_0$:
\begin{subequations} \label{eq:Tt_Jordan}
\begin{gather}
L_0T = 2T, \\
L_0t = 2t+T, \\
\overbar L_0T = 0, \qq{and} \\
\overbar L_0t = T.
\end{gather}
\end{subequations}
For the two-point functions involving $T$ and $t$, we have (at $c = 0$)
\begin{subequations}
\begin{gather}
\ev*{T(z)T(0)} = 0, \\
\ev*{T(z)t(0)} = \frac{b}{z^4}, \qq{and} \\
\ev*{t(z)t(0)} = \frac{-2b\log z+\theta}{z^4},
\end{gather}
\end{subequations}
where $\theta$ is an arbitrary constant that can be adjusted using the freedom $t\to t+\lambda T$, with $\lambda$ an arbitrary scalar. Recalling the Virasoro inner product (for holomorphic fields) \cite{DFMS1997}
\begin{equation}
\braket*{\Phi}{\Psi} = z^{2h_\Phi}\lim_{z\to\infty}\ev*{\Phi(z)\Psi(0)},
\end{equation}
we have, in the CFT, $\braket*{T} = 0$, $\braket*{t} = \infty$, and $\braket*{T}{t} = b$.

Since in this case the lattice model exhibits, in finite size, a Jordan block for the states corresponding to $T$ and $t$ as well, the measurement of $b$ would thus seem straightforward---identify $T$ and $t$, and evaluate their conformal scalar product. The difficulty with this approach is the otherwise-mundane matter of normalization. Because $\ip*{T} = 0$ in the CFT, also observed in finite size on the lattice, the value of $\ip*{T}{t}$ on the lattice is completely dependent on the way one normalizes $T$ (note that we will in general use the same symbol to denote states in the continuum limit and on the lattice: the context should make the distinction clear). The solution is to write an expression for $b$ that is independent of the normalization of $T$. In the CFT, $T = L_{-2}I$, with $I$ the identity field. Being primary, we may normalize its lattice analogue to unity: $\ip*{I} = 1$. Then
\begin{equation}\label{bmeasure}
b = \frac{b^2}{b} = \frac{|\mel*{t}{L_{-2}}{I}|^2}{\ip*{t}{T}}
\end{equation}
is independent of the normalization of $T$, since the normalization of $t$ is linked to that of $T$ via the coefficients in \eqref{eq:Tt_Jordan}. Calculating the lattice equivalent of Eq.~\eqref{bmeasure} is well known to give results converging to the expected value of $b$ as $L$ increases.

The method of measuring $b$ between $T$ and $t$ generalizes to any logarithmic pair, especially regarding the fact that $T$ is a null state. In general, whenever a Jordan block forms, there is a top and a bottom state. If this Jordan block comes from a Hamiltonian describing a logarithmic CFT, from general arguments involving the action of the Virasoro algebra, it is expected that the bottom state in such a Jordan block has zero conformal norm square. In the CFT, calling $\tilde\Psi$ the top state and $\Psi$ the bottom state, the result follows from the fact that we have 
\begin{subequations}
\begin{gather}
L_0\tilde\Psi = h\tilde\Psi + \Psi \qq{and}\\
L_0\Psi = h\Psi.
\end{gather}
\end{subequations}
Hence we have 
\begin{equation}
\ip*{\Psi}=\ip*{\Psi}{(L_0-h)\tilde\Psi}=\ip*{(L_0-h)\Psi}{\tilde\Psi}=0
\end{equation}
where in the last equation we used the fact that $L_n^\dagger=L_{-n}$ (i.e. $L_n$ and $L_{-n}$ are conjugate for the Virasoro norm).

For the problem at hand, we only observe pairs of singlets on the lattice: they have close eigenvalues in finite size that approach each other in the limit $L\to\infty$. Nevertheless, both singlets are still proper eigenvectors, and we expect that they both converge to the bottom state $\Psi$ of an emerging Jordan block. They should thus have a loop norm going to zero in the limit $L\to\infty$. This will be checked in cases where we find emerging Jordan blocks using the $J$ measure.

The method for calculating $b$ in this case---generalizing the procedure described around Eq.~\eqref{bmeasure} when there is a Jordan block is finite size---will be as follows:
\begin{enumerate}
\item Identify an emerging Jordan block by finding pairs of vectors $\Psi$, $\Psi'$ on the lattice corresponding to given conformal fields such that $\lim J(\Psi,\Psi') = 1$.

\item Construct the emerging Jordan vector $\tilde\Psi$ by orthogonalizing $\Psi'$ against $\Psi$ with respect to the standard inner product $(\cdot|\cdot)$. Normalize $\tilde\Psi$ so that $(\Psi|H^{(L)}_0|\tilde\Psi) = 2$ (recall generally that $H^{(L)}_n = \KSL_n + \KSLb_{-n}$).

\item Using the CFT, express $\Psi = A\Phi$ as a descendant of a primary field $\Phi$, with $A$ the corresponding null vector operator. Identify the lattice analogue of $\Phi$, and normalize it to $\ip*{\Phi} = 1$.

\item Finally,
\begin{equation} \label{eq:b_measurement}
b (L)= \frac{|\mel*{\tilde\Psi}{A^{(L)}}{\Phi}|^2}{\ip*{\tilde\Psi}{\Psi}},
\end{equation}
where we note that convergence as $L\to\infty$ may be improved by using $H^{(L)}_{-n}$ instead of $\KSL_{-n}$ in the expression for $A^{(L)}$ (of course this would not change anything in the continuum limit).
\end{enumerate}
There are many possibilities for the unspecified limit in step 1. In this paper we focus on the Jordan blocks that form as $L \to \infty$ by studying the emerging Jordan blocks at finite $L$; we will also use the emerging Jordan blocks formalism to determine (in the appendix) the $b$ number in the well-studied case of the Jordan block between $T$ and $t$ as an illustrative, analytically tractable example that validates the formalism and other observations to be made. 

Additionally, the procedure gives two possible measurements of $b$. Since $\Psi$ and $\Psi'$ both converge to the same vector, they should be treated on an equal footing, and by switching the roles of $\Psi$ and $\Psi'$ we obtain a second value for $b$. Because the value of $b$ should reflect an intrinsic property of the module, it should be insensitive to the method of calculation. The outcome of the procedure is thus only meaningful when the two calculations agree. We will return to this remark once we arrive at concrete results, but we note first that in each instance where we carry out these measurements, the two $b$ numbers agree (within a margin where only numerical computations are possible) precisely where Jordan blocks are known or suspected to appear, and differ otherwise.

\section{Measurements of indecomposability parameters} \label{measurements}

\subsection{Emerging Jordan blocks in $\mathscr W_{1,1}$ at generic $c$}

In $\mathscr W_{1,1}$ there are two eigenstates at zero momentum\footnote{In fact, some states exhibit a finite macroscopic momentum equal to $\pi$ added to their conformal part (which scales like $1/L$)---we do not make the distinction, and only refer in what follows to the conformal part \cite{GrimmSchutz}.} whose conformal weights go to $h + \overbar h = 2$ as $L \to \infty$. For generic $c$ and finite $L$, these eigenstates are non-degenerate, and we refer to them by the labels $\alpha$ and $\beta$. In practice, $\alpha$ is the first excitation (above the ground state) in the zero-momentum sector of $\mathscr W_{1,1}$. $\beta$ is the second or third excitation, depending on the values of $L$ and $c$: at values of $c$ close to 1 and small values of $L$, it is the third excitation, but crosses over to the second excitation at large values of $L$.\footnote{In Table 13 of \cite{Grans-Samuelsson2020}, they are lines $i_{13} = 5$ and $6$. Cf.\ end of Section \ref{field_identification}.}

\begin{figure}
\centering
\includegraphics[width=0.8\textwidth]{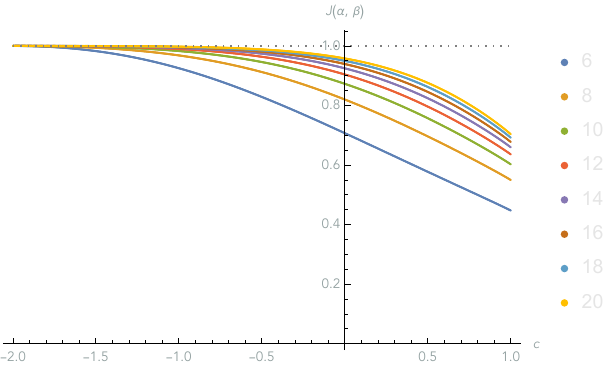}
\caption{The formation of a Jordan block between two singlets in $\mathscr W_{1,1}$. The legend indicates the value of $N = 2L$. It is expected that the curves extrapolate to the dashed line in the limit $L\to\infty$.}
\label{W11J2}
\end{figure}

The emerging Jordan block involving $\alpha$ and $\beta$, based on the measure $J(\alpha,\beta)$, can be seen in Figure \ref{W11J2}. The plot\footnote{All data is collected by subdividing the interval $(-2, 1)$ representing the range of $c$ considered in this paper into even divisions of width $\pi/1000$, so that measurements are carried out for $c \in \{-636\pi/1000, -635\pi/1000,\ldots,318\pi/1000\}$. The point $c = 0$ is omitted where consequences of it being non-generic are manifest---i.e., in the glued module $\overbar{\mathscr N}_{\!\!2}$.} suggests that there is a Jordan block at finite size for $c = -2$, and this is indeed the case \cite{GRS2013a}.

\begin{figure}
\centering
\includegraphics[width=0.8\textwidth]{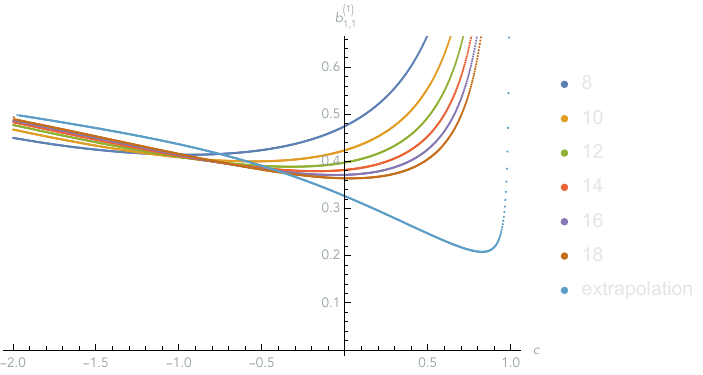}
\caption{The measurement of $b_{1,1}^{(1)}$ between $\alpha$ and $\beta$, together with a polynomial extrapolation.}
\label{W11b1}
\end{figure}

\begin{figure}
\centering
\includegraphics[width=0.8\textwidth]{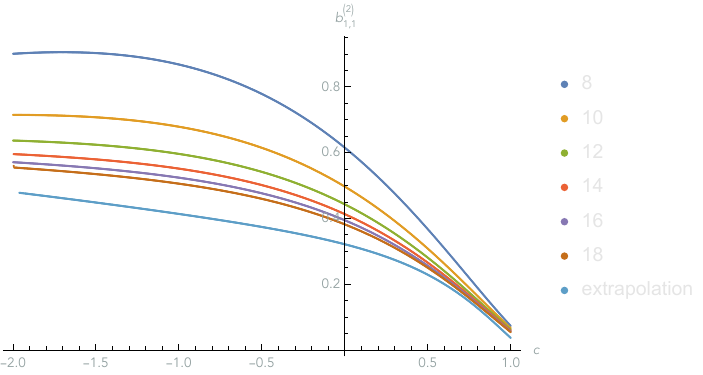}
\caption{The measurement of $b_{1,1}^{(2)}$ between $\alpha$ and $\beta$, together with a polynomial extrapolation.}
\label{W11b2}
\end{figure}

Following our procedure, we calculate the parameter $b_{1,1}$ that describes this Jordan block at finite size. Both $\alpha$ and $\beta$ converge to the same vector as $L\to\infty$, the primary field $\Psi_{1,1} = L_{-1}\Phi_{1,1} = \overbar L_{-1}\overbar\Phi_{1,1}$ of Eq.~\eqref{eq:indecomposable_diamond} (again, the equality holds in the continuum limit). Take, for the lattice approximation of $\Phi_{1,1}$, the lowest field\footnote{This field is line $i_{13} = 2$ in Table 13 of \cite{Grans-Samuelsson2020}.} of momentum 1 and normalize it to $\ip*{\Phi_{1,1}} = 1$. Then, orthogonalize $\alpha$ and $\beta$ by subtracting from $\beta$ its component along $\alpha$, resulting in $\tilde\beta$, and normalize it such that $(\alpha|H_0|\tilde\beta) = 2$. Then, we obtain a lattice measurement of the logarithmic coupling by calculating
\begin{equation}
b^{(1)}_{1,1}(L,c) = \frac{|\mel*{\tilde\beta}{H^{(L)}_{-1}}{\Phi_{1,1}}|^2}{\ip*{\tilde\beta}{\alpha}}.
\end{equation}
As $\alpha$ and $\beta$ should be treated on an equal footing, we similarly define $b^{(2)}_{1,1}$ by exchanging the roles of $\alpha$ and $\beta$ in this procedure. $b^{(2)}_{1,1}$ will generically be different from $b^{(1)}_{1,1}$, but when a genuine Jordan block forms, they should be equal, as we see in the analytic examples. These two measurements of $b_{1,1}$ are given graphically in Figures \ref{W11b1} and \ref{W11b2}.


\begin{figure}
\centering
\includegraphics[width=0.8\textwidth]{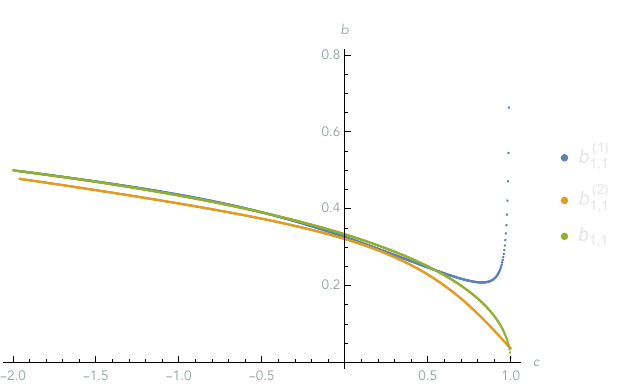}
\caption{The extrapolations of $b_{1,1}$ compared with the expected value.}
\label{W11bs}
\end{figure}

We predicted in Eq.~\eqref{eq:b11} that
\begin{equation}
b_{1,1} = \frac{1}{x+1}.
\end{equation}
The comparison of this value with the extrapolations is given in Figure \ref{W11bs}.

\subsection{Emerging Jordan blocks in $\mathscr W_{2,1}$ at generic $c$}
In $\mathscr W_{2,1}$ there are four fields that have $h + \overbar h = 2h_{1,-2}$. For generic $c$ and finite $L$, these fields correspond to two non-degenerate eigenvalues (singlets), which we denote by $\mu$ and $\nu$, and one doubly degenerate eigenvalue. These four fields tend to have weights higher than $\Phi_{0,2}$ and $L_{-1}\overbar L_{-1}\Phi_{0,2}$.\footnote{In Table 14 of \cite{Grans-Samuelsson2020}, we identify the singlets as lines $i_{13} = 24$ and $35$, and the doublet as line $i_{13} = 25$.} At generic $c$, the two singlets are expected to form a Jordan block in the $L\to\infty$ limit. The doublet occurs because of a left--right (chiral--antichiral) symmetry \cite{Grans-Samuelsson2020}, and is not expected to be involved in the Jordan-block structure.

\begin{figure}
\centering
\includegraphics[width=0.8\textwidth]{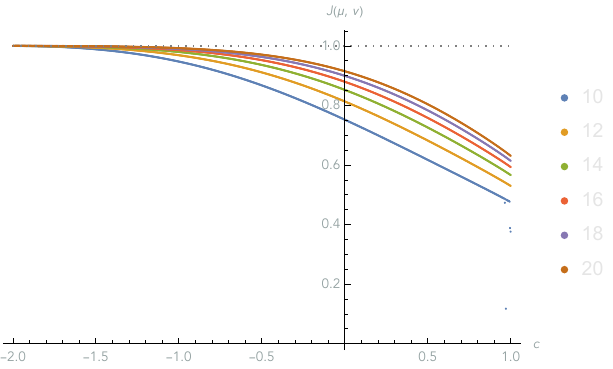}
\caption{The formation of a Jordan block between two singlets in $\mathscr W_{2,1}$. The legend indicates the value of $N = 2L$.}
\label{W21J4}
\end{figure}

Using the $J$ measure, we can directly observe the emerging Jordan block involving $\mu$ and $\nu$ (Figure \ref{W21J4}). As with the fields $\alpha$ and $\beta$ in $\mathscr W_{1,1}$, the plot correctly suggests that there is a Jordan block at finite size for $c = -2$.

\begin{figure}
\centering
\includegraphics[width=0.8\textwidth]{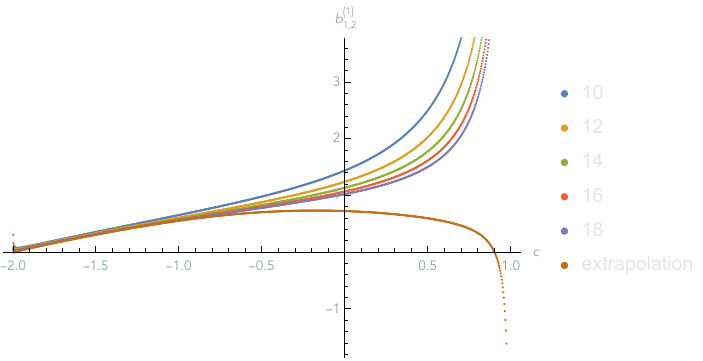}
\caption{The measurement of $b_{1,2}^{(1)}$ between $\mu$ and $\nu$.}
\label{W21b1}
\end{figure}

\begin{figure}
\centering
\includegraphics[width=0.8\textwidth]{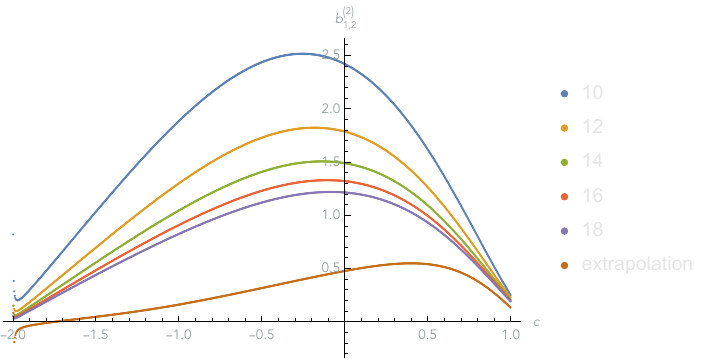}
\caption{The measurement of $b_{1,2}^{(2)}$ between $\mu$ and $\nu$.}
\label{W21b2}
\end{figure}

We again obtain estimates for the parameter $b_{1,2}$ that characterizes the Jordan block involving $\mu$ and $\nu$. Both $\mu$ and $\nu$ converge to the same vector as $L\to\infty$, the field $\Psi_{1,2}$ in Eq.~\eqref{eq:indecomposable_diamond}. The field $\Phi_{1,2}$ is realized on the lattice as the lowest field\footnote{This field is line $i_{13} = 3$ in Table 14 of \cite{Grans-Samuelsson2020}.} of momentum $2$, which we normalize to $\ip*{\Phi_{1,2}} = 1$. Orthogonalize $\mu$ and $\nu$ by subtracting from $\nu$ its component along $\mu$, resulting in $\tilde\nu$, and normalize it so that $(\mu|H_0|\tilde\nu) = 2$. Then we have
\begin{equation}
b^{(1)}_{1,2}(L,c) = \frac{|\mel*{\tilde\nu}{A^{(L)}}{\Phi_{1,2}}|^2}{\ip*{\tilde\nu}{\mu}},
\end{equation}
where
\begin{equation}
A^{(L)} = H^{(L)}_{-2} - \frac{3}{2(2h_{1,2} + 1)}(H^{(L)}_{-1})^2,
\end{equation}
as in the continuum version \eqref{A_12}. We similarly define $b^{(2)}_{1,2}$ by exchanging the roles of $\mu$ and $\nu$ in this procedure. The two measurements of $b_{1,2}$ are given graphically in Figures \ref{W21b1} and \ref{W21b2}.


\begin{figure}
\centering
\includegraphics[width=0.8\textwidth]{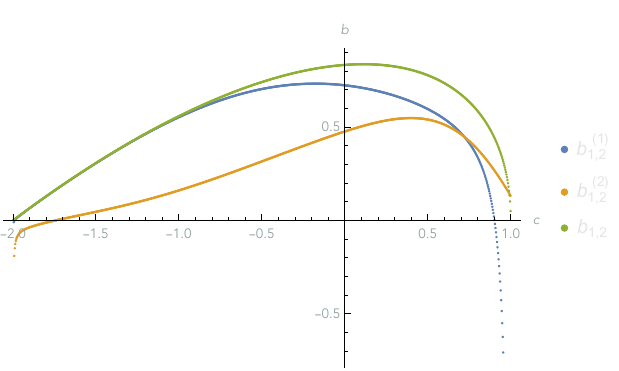}
\caption{The two extrapolations of $b_{1,2}$ compared with the expected value.}
\label{W21bs}
\end{figure}

From Eq.~\eqref{eq:b12} we have
\begin{equation}
b_{1,2} = \frac{4}{x+1} - \frac{2}{x^2}.
\end{equation}
The comparison of this value with the two extrapolated measurements is shown in Figure \ref{W21bs}.

\sectionbreak

Particularly with the measured values of $b_{1,1}^{(1)}$ and $b_{1,2}^{(1)}$, we see excellent agreement with the theoretical $b$ values in Eqs.~\eqref{eq:b11} and \eqref{eq:b12}. As these values of $b$ were derived from the diamond structure of Eq.~\eqref{eq:indecomposable_diamond}, these measurements almost certainly establish the correctness of Conjecture \ref{loop_j}. Admittedly, the measurement is not as compelling near $c = 1$, and the values of $b_{1,1}^{(1)}$ and $b_{1,2}^{(1)}$ differ significantly from $b_{1,1}^{(2)}$ and $b_{1,2}^{(2)}$, even though they should give the same values in the limit. We remarked previously that the $b$ number should be a characteristic of the module, insensitive to the method of calculation. In the analytical examples we are able to take limits exactly, and we see an agreement precisely where the two $b$ values are equal with the existence of a Jordan block. We believe the discrepancies in the pairs of numerical measurements are a result of the limited amount of accessible data and the necessity of extrapolation, which certainly introduces some errors. With further data and improved extrapolations, we would expect the discrepancies to shrink and the two extrapolated $b$ values to overlap with the theoretical curve, thereby establishing the existence of the Jordan block only in the continuum limit. We note additionally that the emergence of the Jordan block is much slower near $c = 1$, as seen in Figures \ref{W11J2} and \ref{W21J4}, and this may be remedied too once we have access to larger lattice sizes.

\bigskip
We note that it is possible to carry out the procedures of the preceding two subsections in the glued modules $\overbar{\mathscr N}_{\!\!1}$ or $\overbar{\mathscr N}_{\!\!2}$ instead of the standard modules as described above. The numerical results---in particular, the $J$ measure and $b$ values---in finite size are slightly different, as the eigenvectors of the Hamiltonian contain components with various numbers of through-lines and the inner product is modified, but they are compatible with the same expected limit as $L\to\infty$.

\section{Conclusion} \label{conclusion}

We have shown in this paper how to study indecomposable Virasoro modules starting from a lattice model with diagonalizable Hamiltonian using the concept of emerging Jordan blocks. This allowed us to confirm the expected logarithmic properties of the CFTs for the $Q$-state Potts model (as well as the low-temperature $O(\m)$ model). The next step in this analysis would be to investigate systematically the case $c=0$, and confirm in particular the existence of a rank-3 Jordan block for fields with conformal weights $(h,\bar{h})=(2,2)$ as conjectured in \cite{HeSaleur2022}. We will leave this (highly technical) exercise for a subsequent work.

\bigskip
\noindent {\bf Acknowledgments:} we thank L.\ Grans-Samuelsson, Y.\ He, R.\ Nivesvivat, and S.\ Ribault for discussions and related collaborations. This work was supported in part by the French Agence Nationale de la Recherche (ANR) under grant ANR-21-CE40-0003 (project CONFICA). Parts of the paper build upon work contained in a thesis submitted by L.L.\ in partial fulfillment of the requirements for a doctoral degree at University of Southern California \cite{Liu2024}.
\appendix
\section{Other results obtained with the emerging Jordan blocks method}

The results of this paper rely on the new technique of studying conjectured Jordan blocks via a sequence of operators that have only trivial Jordan blocks---i.e., what we have called the emerging Jordan blocks method. Our indirect method of measuring $b_{1,1}$ and $b_{1,2}$ can therefore be better justified if we can use this technique to replicate results obtained in situations where Jordan blocks have actually been observed. 

The most well-studied of these is the Jordan block between the stress--energy tensor $T$ and its logarithmic partner $t$ at $c = 0$. In Section \ref{sec:absence}, we describe how this Jordan block can be observed in a finite-dimensional model. In Section \ref{sec:bTt}, we use the emerging Jordan blocks technique to compute the $b$ number characterizing this Jordan block, with the additional findings of some closed-form expressions for finite-size $b$ measurements. In short, we find with the emerging Jordan blocks technique the same values present in the literature computed with Jordan blocks, thus giving credence to the method.

\subsection{Absence and restoration of Jordan blocks} \label{sec:absence}
\subsubsection{The contraction parameter $y$}

The modules we study in this paper are $\mathscr W_{1,1}$, $\mathscr W_{2,1}$, and $\overbar{\mathscr N}_{\!\!2} = \overbar{\mathscr W}_{\!\!0,\q^{\pm 2}} + \mathscr W_{1,1} + \mathscr W_{2,1}$. As mentioned, in finite size Jordan blocks are not observed in $\mathscr W_{1,1}$ and $\mathscr W_{2,1}$. Technically, this situation occurs because the fields expected to form such a Jordan block are found with slightly different eigenvalues, precluding that possibility. The Jordan blocks emerge as the pairs of eigenvalues converge in the continuum limit, and, as will be seen, this situation can be studied successfully with the ideas in Section \ref{strategy}.

Strictly at $c = 0$ ($\m = 1$), it so happens that there are no Jordan blocks on the lattice for the dense loop model Hamiltonian, even in the glued module $\overbar{\mathscr N}_{\!\!2}$. This is a somewhat surprising fact since we know from representation theory that the JTL algebra is not semisimple at that point, and that the modules $\overbar{\mathscr W}_{\!\!0,\q^{\pm 2}}$ and $\mathscr W_{2,1}$ are glued by the action of generic elements of the algebra. Furthermore, other representations, such as the $s\ell(2|1)$ spin chain, do exhibit the expected Jordan blocks. Nonetheless, the fact remains that the Hamiltonian itself is fully diagonalizable, and that, for instance, the eigenvalues corresponding to the conformal states $T$ and its logarithmic partner $t$ are degenerate without being part of a Jordan block. This fact was observed previously by \cite{DJS2010c} for the case of open boundary conditions, though it was not clear at the time whether this was a bizarre effect due to the small sizes studied. However, the situation does not seem to change when exploring larger sizes and with periodic boundary conditions, and it seems to be a definite feature of the model.

In the preceding work as well as in subsequent studies, this problem was circumvented by introducing a parameter $y\in\mathbb R$ that did not change the (generalized) eigenvalues of the Hamiltonian but allowed the Jordan blocks at $c = 0$ to ``reappear'' somewhat miraculously. Whenever certain parity conditions to be specified were met, the result would be multiplied by a factor $y$. The parity conditions are as follows: whenever two through-lines become contracted, and the ``left'' one is on an odd\footnote{Whether the ``special'' sites are chosen to be odd or even makes no difference in the results.} site, then the factor $y$ is applied. For example, consider the action of the diagram
\begin{equation}
a = \vcenter{\hbox{\begin{tikzpicture}
 \newcommand{\dist}{0.2}
 \begin{scope}[yscale=-1,xscale=1]
 	\draw[thick] (2*\dist,0) arc (-180:0:\dist);
	\draw[thick] (0,0) arc (-180:0:3*\dist);
	\draw[thick] (8*\dist,0) -- (8*\dist,-3*\dist);
	\draw[thick] (10*\dist,0) -- (10*\dist,-3*\dist);
 \begin{scope}[xshift=0cm,yshift=-5*\dist cm]
 \begin{scope}[yscale=-1,xscale=1]
 	\draw[thick] (4*\dist,0) arc (-180:0:\dist);
	\draw[thick] (0,0) arc (-180:0:\dist);
	\draw[thick] (8*\dist,0) -- (8*\dist,-3*\dist);
	\draw[thick] (10*\dist,0) -- (10*\dist,-3*\dist);
 \end{scope}
 \end{scope}
 \end{scope}
\draw[thick,dotted] ($(current bounding box.north east) + (0.05+\dist,0.05)$) rectangle ($(current bounding box.south west)+ (-0.05-\dist,-0.05)$);
 \end{tikzpicture}}}
\end{equation}
on the state $\lambda = (1)(2)(3)(4)(56)$. When $a$ is stacked on top of $\lambda$, the curves in the lower row of $a$ contract two pairs of through-lines in the link state $\lambda$: the pair 1--4 and the pair 2--3. The left sites of each pair are $1$ and $2$. The factor $y$ is thus applied once, since only $1$ is odd. The result of the computation is thus $a\lambda = y(12)(34)(56)$. The same contracted pair can give a factor of $y$ or not depending on which one sits on the left: for $N = 2$, $e_1(1)(2) = y(12)$ but $e_2(1)(2) = (21)$, since in the latter case site $2$ sits on the ``left,'' connected to $1$ through the periodic boundary (note that any contracted pair of through-lines will be on sites of opposite parity). The base model is thus recovered when $y = 1$. The meaning of this parameter was not very clear, except for the fact that it broke the symmetry of translation by one site---a somewhat pleasant feature, since the dense loop model is naturally described using an underlying oriented lattice that is only compatible with translation invariance by two sites. With $y \ne 1$ for open boundary conditions ($y \ne \pm 1$ for periodic boundary conditions) the measured values of the logarithmic couplings known at the time of those studies for $c = 0$ were found to be independent of $y$ and in excellent agreement with theoretical expectations.

The parameter $y$ only affects matrix elements between different modules. As long as one restricts to a single standard module (e.g., $\mathscr W_{1,1}$ or $\mathscr W_{2,1}$), no calculations can be affected by the value of $y$---and this is true even at $c = 0$. However, it turns out that, if we measure the quantities $J$, $b_{1,1}$, and $b_{1,2}$ by carrying out corresponding calculations in the glued module, their values depend on $y$ in a nontrivial, most unsatisfactory way: they encounter seemingly-random divergences and sign changes when considered as functions of $c$, and these divergences and sign flips change positions as $y$ varies. This occurs despite the fact that the JTL algebra is simple for $c \ne 0$, so a change of basis provides modules isomorphic to $\mathscr W_{1,1}$ or $\mathscr W_{2,1}$. The only strategy to recover results which are compatible with conformal invariance seems to be to consider the limit $y \to 1$ at finite $L$, and then extrapolate to the thermodynamic limit, if it can be done. It is fair to ask why one would have to go through these gymnastics, and what they mean.

\subsubsection{Symmetry of the JTL algebra}

We now observe that the JTL algebra has a symmetry under $e_j\to -e_j$ and $\m\to -\m$: more precisely, if we set $e'_j\equiv-e_j$, the $e'_j$ satisfy the JTL relations with $\m'=-\m$. Note that because of the relation
\begin{equation}
 \tau^2e_{N-1}=e_1\cdots e_{N-1}
\end{equation}
this is only true for $N$ even. Note also that, using the XXZ representation of the Temperley--Lieb algebra 
\begin{equation}
 e_j=-\sigma_j^-\sigma_{j+1}^+-\sigma_j^+\sigma_{j+1}^--\frac{\cos\gamma}{2}\sigma_j^z\sigma_{j+1}^z -\frac{\i\sin\gamma}{2}(\sigma_j^z-\sigma_{j+1}^z)+\frac{\cos\gamma}{2}
 \end{equation}
this mapping corresponds to $\sigma_j^{x,y}\to (-1)^j\sigma_{j}^{x,y}$ and $\gamma\to \gamma-\pi$, which leads, in particular, to the famous mapping between the XXZ Hamiltonian $H_{XXZ}(\Delta)$
 and $-H_{XXZ}(-\Delta)$ for an even number of sites\footnote{We note that the situation would be more complicated for an odd number of sites. This case is not relevant for the present paper---see \cite{Stroganov2001} for other features particular to odd $N$.}, where 
\begin{equation}
H_{XXZ}=\sum_{j=1}^N (\sigma_j^x\sigma_{j+1}^x+\sigma_j^y\sigma_{j+1}^y+\Delta \sigma_j^z\sigma_{j+1}^z).
\end{equation}

We can therefore study the loop model with opposite Hamiltonian and opposite value of the loop fugacity. While it is widely expected that these two descriptions are equivalent, they do differ right at $\m=1$. To see this consider the simple example of $2L = 4$. Using the basis $\{(12)(34),(23)(14),(2)(3)(41),(3)(4)(12),(1)(4)(23),(1)(2)(34),(1)(2)(3)(4)\}$, we have
\begin{equation}
H_4(\m,e_\infty) = \begin{pmatrix}
 4 e_\infty-2\m & -2 & 0 & -1 & 0 & -1 & 0 \\
 -2 & 4 e_\infty-2\m & -1 & 0 & -1 & 0 & 0 \\
 0 & 0 & 4 e_\infty-\m & -1 & 0 & -1 & -1 \\
 0 & 0 & -1 & 4 e_\infty-\m & -1 & 0 & -1 \\
 0 & 0 & 0 & -1 & 4 e_\infty-\m & -1 & -1 \\
 0 & 0 & -1 & 0 & -1 & 4 e_\infty-\m & -1 \\
 0 & 0 & 0 & 0 & 0 & 0 & 4 e_\infty
\end{pmatrix}.
\end{equation}
Both 
\begin{equation}
H_4(1,1) = \begin{pmatrix}
 2 & -2 & 0 & -1 & 0 & -1 & 0 \\
 -2 & 2 & -1 & 0 & -1 & 0 & 0 \\
 0 & 0 & 3 & -1 & 0 & -1 & -1 \\
 0 & 0 & -1 & 3 & -1 & 0 & -1 \\
 0 & 0 & 0 & -1 & 3 & -1 & -1 \\
 0 & 0 & -1 & 0 & -1 & 3 & -1 \\
 0 & 0 & 0 & 0 & 0 & 0 & 4
\end{pmatrix}
\end{equation} and 
\begin{equation}
-H_4(-1,-1) = \begin{pmatrix}
 2 & 2 & 0 & 1 & 0 & 1 & 0 \\
 2 & 2 & 1 & 0 & 1 & 0 & 0 \\
 0 & 0 & 3 & 1 & 0 & 1 & 1 \\
 0 & 0 & 1 & 3 & 1 & 0 & 1 \\
 0 & 0 & 0 & 1 & 3 & 1 & 1 \\
 0 & 0 & 1 & 0 & 1 & 3 & 1 \\
 0 & 0 & 0 & 0 & 0 & 0 & 4 
\end{pmatrix}
\end{equation} have eigenvalues $\{\lambda_1,\ldots,\lambda_7\} = \{0,1,3,3,4,4,5\}$. It turns out however that $H_4(1,1)$ is diagonalizable with the 7 corresponding geometric eigenvectors
\begin{subequations}
\begin{gather}
v_1 = (1, 1, 0, 0, 0, 0, 0), \\
v_2 = (2, 2, -1, -1, -1, -1, 0), \\
v_3 = (0, 0, 1, 0, -1, 0, 0), \\
v_4 = (0, 0, 0, 1, 0, -1, 0), \\
v_5 = (1, -1, 0, 0, 0, 0, 0), \\
v_6 = (1, 0, -1, -1, -1, -1, 3), \\
v_7 = (2, -2, 1, -1, 1, -1, 0),
\end{gather}
\end{subequations}
but $-H_4(-1,-1)$ is not. Its generalized eigenbasis is
\begin{subequations} \label{eq:eigenvectors_m_-1}
\begin{gather}
v_1 = (1, -1, 0, 0, 0, 0, 0), \\
v_2 = (2, -2, -1, 1, -1, 1, 0), \\
v_3 = (0, 0, 0, 1, 0, -1, 0), \\
v_4 = (0, 0, 1, 0, -1, 0, 0), \\
v_5 = (1, 1, 0, 0, 0, 0, 0), \\
\tilde v_6 = \left(0, 0, \frac{1}{2}, \frac{1}{2}, \frac{1}{2}, \frac{1}{2}, -\frac{1}{2}\right), \\
v_7 = (2, 2, 1, 1, 1, 1, 0),
\end{gather}
\end{subequations}
where the tilde on $\tilde v_6$ indicates it is not a proper eigenvector.

We have now recovered nontrivial Jordan blocks without invoking the parameter $y$. Let us reinforce this observation using the formalism of emerging Jordan blocks. With $\m$ remaining arbitrary, we have more generally the corresponding eigenvectors (of $H(\m,e_\infty)$ and $-H(-\m,-e_\infty)$, where $\m > 0$)
\begin{subequations} \label{eq:eigenvectors_generic_m}
\begin{gather}
u_5 = (1, -1, 0, 0, 0, 0, 0), \\
u_6 = (1, 1, -(\m+1), -(\m+1), -(\m+1), -(\m+1), (\m+1)(\m+2)), \\
v_5 = (1, 1, 0, 0, 0, 0, 0), \\
v_6 = (1, 1, \m-1, \m-1, \m-1, \m-1, (\m-1)(\m-2)).
\end{gather}
\end{subequations}
One can immediately see that $J(u_5, u_6) = 0$, while $\lim_{\m\to 1}J(v_5, v_6) = 1$. More generally,
\begin{equation}
J(v_5, v_6) = \sqrt{\frac{2}{2 + 4(\m-1)^2 + (\m-1)^2(\m-2)^2}}.
\end{equation}
Furthermore, $\tilde v_6$ is obtained by orthogonalizing $v_6$ against $v_5$, then rescaling the result in the limit $\m \to 1$. We conclude the presentation of our method on this example with the measurement of $b$ in the next section.

Because the modified model with $\m \to -\m$ and $e_j \to -e_j$ shows the expected structure at $\m=1$, in agreement with other representations, we will consider it to be the ``correct'' loop model. Hereafter, we continue to use the parametrization $x \in (0,\infty)$, but now with
\begin{subequations}
 \begin{gather}
 \m = -2\cos\left(\frac{\pi}{x+1}\right), \\
 e_j = (-1)\times{}\raisebox{-5mm}{
\begin{tikzpicture}[scale=0.6]
 \draw[thick] (0,-1)--(0,1) node[above]{\scriptsize $1$};
\end{tikzpicture}}
\quad\cdots\quad
\raisebox{-5mm}{\begin{tikzpicture}[scale=0.6]
 \draw[thick] (-1,-1)--(-1,1);
 \draw[thick] (2,-1)--(2,1);
 \draw[thick] (0,-1)--(0,-0.5) arc(180:0:5mm and 4mm)--(1,-1);
 \draw[thick] (0,1) node[above]{\scriptsize $j$}--(0,0.5) arc(180:360:5mm and 4mm)--(1,1) node[above]{\scriptsize $j+1$};
\end{tikzpicture}}
\quad\cdots\quad
\raisebox{-5mm}{\begin{tikzpicture}[scale=0.6]
 \draw[thick] (0,-1)--(0,1) node[above]{\scriptsize $N$};
\end{tikzpicture}}
 \end{gather}
\end{subequations}
The expressions for the Hamiltonian and Koo--Saleur generators remain the same, evaluated using the new values of $\m$ and $e_j$. The spectra of the two Hamiltonians are identical, so that the identification of fields is not affected.

\subsection{The parameter $b$ between $T$ and $t$} \label{sec:bTt}

Although we now have real, rather than emerging, Jordan blocks, it is useful to consider the measurement of $b$ using emerging Jordan blocks anyway, because we already have measurements of $b$ itself using the Jordan blocks \cite{VGJS2012}, and we would like to see agreement between the two methods. The procedure of Section \ref{b_emerging} reads as follows. For fixed $L$ and generic $c$ close to 0, identify an eigenvector of $H$ in the representation $\overbar{\mathscr W}_{\!\!0,\mathfrak q^{\pm 2}} + \mathscr W_{1,1} + \mathscr W_{2,1}$ corresponding to the stress--energy tensor $T$. (Actually, it is known that $T$ lives in $\overbar{\mathscr W}_{\!\!0,\mathfrak q^{\pm 2}}$, and since the action of JTL can only decrease the first index $j$, it suffices to find $T$ in $\overbar{\mathscr W}_{\!\!0,\mathfrak q^{\pm 2}}$ and embed it in the full glued module.) Find an eigenvector $T'$ that becomes degenerate and parallel with $T$ at $c = 0$, but is otherwise distinct---it has nonzero components in $\mathscr W_{2,1}$. Define $t$ via the Gram--Schmidt process and normalize it as specified. Then
\begin{equation}
b(L) = \lim_{c\to 0} \frac{|\mel*{t}{H_{-2}}{I}|^2}{\ip*{t}{T}\ip*{I}{I}},
\end{equation}
where $I$ is the ground state (fully contained in $\overbar{\mathscr W}_{\!\!0,\mathfrak q^{\pm 2}}$) and we have taken $A = L_{-2}$ as $T = L_{-2}I$.

The last expression requires some explanation. The modifications $e_j \to -e_j$ and $\m \to -\m$ must be propagated throughout the entire process. This includes not only the Hamiltonian but also the inner product and Koo--Saleur generators. When we do this, it is found that $\ip*{I} = (-1)^L$ upon normalization---i.e., it is negative for odd $L$. The sign of the norm square is invariant and cannot be changed by rescaling, even by a complex constant. However, because CFT demands that primary fields have unit norm square, particularly the ground state, in these cases we must enforce this by manually adding in the sign, and declaring this to be the correct inner product. This is the meaning of the appearance of $\ip*{I}$ in the denominator.

The results of this process are shown in Table \ref{b_W012}.
\begin{table}[h]
\centering
\begin{tabular}{cc}
\toprule
$N = 2L$ & $b$ \\
\midrule
4 & $-1.96028$ \\
6 & $-3.24046$ \\
8 & $-3.94952$ \\
10 & $-4.33295$ \\
12 & $-4.55079$ \\
14 & $-4.68234$ \\
16 & $-4.76633$ \\
18 & $-4.82253$ \\
\midrule
conjectured & $-5$ \\
\bottomrule
\end{tabular}
\caption{Different measurements for $b = \ip*{t}{T}$ in $\overbar{\mathscr W}_{\!\!0,\mathfrak q^{\pm 2}} + \mathscr W_{1,1} + \mathscr W_{2,1}$.}
\label{b_W012}
\end{table}

In this case, when the roles of $T$ and $T'$ are exchanged, the same value of $b(L)$ is obtained, unlike what is observed in Section \ref{b_emerging}. The two measurements match here because the limit $c \to 0$ is actually reached, rather than extrapolated as must be the case for the limit $L \to \infty$. Furthermore, the measurements match those reported in \cite{VGJS2012}. The crucial difference is that we were able to compute the same quantities without ever calculating a Jordan form, a demanding task. We needed only to diagonalize the Hamiltonian and take limits; the emerging Jordan block took care of the indecomposability parameter.

Because exact diagonalization is also algebraically much less daunting a task than finding a Jordan canonical form, we are able to calculate closed form expressions for $b(L)$ for sufficiently small $L$. For $L = 2$, $T$ and $T'$ are $v_5$ and $v_6$ of Eq.~\eqref{eq:eigenvectors_generic_m}, and $I$ is $v_1/2$ of Eq.~\eqref{eq:eigenvectors_m_-1}, so that it is normalized to $\ip*{I} = 1$. This leads to two measurements of $b$:
\begin{subequations}
\begin{gather}
b^{(1)} = -\frac{8(1-\m)}{\pi v_{\text F}}, \qq{and} \\
b^{(2)} = \frac{4\m(1-\m)(\m^4 - 2\m^3 + \m^2 + 12\m + 10)^2}{\pi v_{\text F} (\m^3 - 7\m^2 + 6\m + 16)}.
\end{gather}
\end{subequations}
When $\m \to -1$, $v_{\text F} \to 3\sqrt 3/2$, and $b^{(1)}, b^{(2)} \to -32\sqrt 3/9\pi \approx -1.96028$, precisely the measured value. An analogous calculation is possible for $L = 3$. One value of $b$ is
\begin{equation}
b^{(1)} = \frac{3}{\pi v_{\text F}}\frac{4 \m^2 + \frac{2\m (2 \m^2-9)}{\sqrt{\m^2+48}}-6}{1+\m}.
\end{equation}
The expression for $b^{(2)}$ is much too complicated to be worth showing, and is not particularly illuminating. However, for $\m \to -1$ both tend to $b = -288\sqrt 3/49\pi \approx -3.24046$, again the measured value. (Despite the denominator in the expression for $b^{(1)}$, it is a removable singularity, and yields a finite value as $\m \to -1$.) Finally, it turns out that at $L = 4$, one may repeat the same exercise to find at $c = 0$
\begin{equation}
b = -\frac{8\sqrt{2105222 + 1531774\sqrt{17}/3}}{1323\pi} \approx -3.94952.
\end{equation}

\uspunctuation
\emergencystretch=1em
\printbibliography
\end{document}